\documentclass[preprint,prd,amsmath,amssymb,nofootinbib,tightenlines,floatfix]{revtex4}
\usepackage[dvips]{graphics}
\usepackage{epsfig}
\usepackage{latexsym}

\newcommand{\lsim}{\buildrel < \over {_\sim}}
\newcommand{\gsim}{\buildrel > \over {_\sim}}

\newcommand{\be}{\begin{equation}}
\newcommand{\ee}{\end{equation}}
\newcommand{\bea}{\begin{eqnarray}}
\newcommand{\eea}{\end{eqnarray}}
\newcommand{\ba}{\begin{array}}
\newcommand{\ea}{\end{array}}

\def\omh{{\omega_{\widetilde H}}} \def\omw{{\omega_{\widetilde W}}}
\def\omb{{\omega_{\widetilde B}}} \def\gamh{{\Gamma_{\widetilde H}}}
\def\gamw{{\Gamma_{\widetilde W}}} \def\gamb{{\Gamma_{\widetilde B}}}
 
\def\kwtilp{{h_{\widetilde W}^+}} \def\khtilp{{h_{\widetilde H}^+}}
\def\nwtilp{{N_{\widetilde W}^+}} \def\nhtilp{{N_{\widetilde H}^+}}

\def\kplp{{h_{pL}^+}}
\def\kprp{{h_{pR}^+}}

\def\khrp{{h_{hR}^+}}

\newcommand{\ket}[1]{\left\lvert #1\right\rangle}
\newcommand{\bra}[1]{\left\langle #1\right\rvert}
\newcommand{\vect}[1]{\mathbf{#1}}
\newcommand{\op}[1]{\textsf{#1}}
\newcommand{\abs}[1]{\left\lvert #1\right\rvert}
\newcommand{\boldgamma}{\mbox{\boldmath$\gamma$}}
\newcommand{\diracslash}[1]{#1\!\!\!/}
\newcommand{\pd}[2]{\frac{\partial #1}{\partial #2}}
\newcommand{\mcdot}{\!\cdot\!}
\newcommand{\CPV}{CP\!\!\!\!\!\!\!\!\raisebox{0pt}{\small$\diagup$}}

\DeclareMathOperator{\Real}{Re}
\DeclareMathOperator{\Imag}{Im}
\DeclareMathOperator{\Tr}{Tr}

\begin{document}

\preprint{Caltech MAP-304}
\preprint{CALT-68-2535}

\title{\Large Resonant Relaxation in Electroweak Baryogenesis} 

\author{\sc Christopher Lee\footnote{
	Electronic address: leec@theory.caltech.edu}}

\author{Vincenzo Cirigliano\footnote{
	Electronic address: vincenzo@caltech.edu}}

\author{Michael J. Ramsey-Musolf\footnote{
	Electronic address: mjrm@krl.caltech.edu}}
\affiliation{California Institute of Technology, Pasadena, CA 91125}

\date{December 23, 2004\\ \vspace{1cm} }

%
\begin{abstract}

We  compute the leading, chiral charge-changing relaxation term in the quantum transport equations that govern electroweak baryogenesis using the closed time path formulation of non-equilibrium quantum field theory. We show that the relaxation transport coefficients may be resonantly enhanced under appropriate conditions on electroweak model parameters and that such enhancements can mitigate the impact of similar enhancements in the $CP$-violating source terms. We also develop a power counting in the time and energy scales entering electroweak baryogenesis and include effects through second order in ratios $\epsilon$ of the small and large scales. We illustrate the implications of the resonantly enhanced ${\cal O}(\epsilon^2)$ terms using the Minimal Supersymmetric Standard Model, focusing on the interplay between the requirements of baryogenesis and constraints obtained from collider studies, precision electroweak data, and electric dipole moment searches.

\end{abstract}

\maketitle


\section{Introduction}
\label{sec:intro}

The origin of the baryon asymmetry of the Universe (BAU) remains an
important, unsolved problem for  particle physics and
cosmology. Assuming that the Universe was matter-antimatter symmetric
at its birth, it is reasonable to suppose that interactions involving
elementary particles generated the BAU during subsequent cosmological
evolution. As noted by Sakharov \cite{Sakharov:1967dj}, obtaining a nonzero BAU requires
both a departure from thermal equilibrium as well as the breakdown of
various discrete symmetries: baryon number ($B$) conservation, charge
conjugation ($C$) invariance, and invariance under the combined $C$
and parity ($P$) transformations\footnote{Allowing for a breakdown of
$CPT$ invariance relaxes the requirement of departure from thermal
equilibrium.}.  The Standard Model (SM) of strong and electroweak
interactions satisfies these conditions and could, in principle,
explain the observed size of the BAU: 
\be
Y_B\equiv \frac{\rho_B}{s} = 
\begin{cases}(7.3\pm 2.5)\times 10^{-11}, & \text{BBN \cite{pdg04}}\\
							(9.2\pm 1.1)\times 10^{-11}, & \text{WMAP \cite{wmap}}
\end{cases}
\ee
where $\rho_B$ is the baryon number density, $s$ is the entropy density of the universe, and where the values shown correspond to 95\% confidence level results obtained from Big Bang Nucleosynthesis (BBN) and the cosmic microwave background (WMAP), respectively. 
In practice, however, neither the strength of the first-order electroweak phase transition in the SM nor the magnitude of SM
$CP$-violating interactions are sufficient to prevent washout of any
net baryon number created by $B$-violating electroweak sphaleron
transitions during the phase transition.

The search for physics beyond the SM is motivated, in part, by the
desire to find new particles whose interactions could overcome the
failure of the SM to explain the BAU. From a phenomenological
standpoint, a particularly attractive possibility is that masses of
such particles are not too different from weak scale
and that their interactions both strengthen the first-order
electroweak phase transition and provide the requisite level of
$CP$-violation needed for the BAU. Precision electroweak measurements
as well as direct searches for new particles at the Tevatron and Large
Hadron Collider may test this possibility, and experiment already
provides rather stringent constraints on some of the most widely
considered extensions of the SM. In the Minimal Supersymmetric Standard
Model (MSSM), for example, present lower bounds on the mass of the lightest
Higgs boson leave open only a small window for a sufficiently strong
first-order phase transition, although this constraint may be relaxed
by introducing new gauge degrees of freedom (see, {\em e.g.}, \cite{Dine:2003ax,Kang:2004pp}). Similarly, limits on the permanent electric dipole moments (EDMs) of elementary
particles and atoms imply that the $CP$-violating phases in the MSSM
must be unnaturally small ($\sim\! 10^{-2})$. Whether such small phases
(supersymmetric or otherwise) can provide for successful electroweak
baryogenesis (EWB) has been an important consideration in past studies
of this problem.

In order to confront phenomenological constraints on the parameters of
various electroweak models with the requirements of EWB, one must
describe the microscopic dynamics of the electroweak phase transition
in a realistic way. Theoretically, the basic mechanism driving
baryogenesis during the phase transition is 
well-established. Weak sphaleron transitions that conserve $B-L$ but
change $B$ and $L$ individually are unsuppressed in regions of
spacetime where electroweak symmetry is unbroken, while they become
exponentially suppressed in regions of broken symmetry. Net baryon
number is captured by expanding regions of broken symmetry (\lq\lq
bubbles"). Given sufficiently strong $C$ and $CP$-violation as well as
departure from thermal equilibrium, the non-zero $B$ generated outside the bubble cannot be entirely washed out by elementary particle interactions that occur at the phase boundary. The baryon number density, $\rho_B$, is governed by a diffusion equation of the form:
\be
\label{eq:rhob1}
{\partial }_t \rho_B(x)  -D\nabla^2\rho_B(x) =
- \Gamma_{\rm ws} F_{\rm ws}(x)[n_L(x) + R\rho_B(x)]\,,
\ee
where $D$ is the diffusion coefficient for baryon number, $\Gamma_{\rm
ws}$ is the weak sphaleron transition rate, $F_{\rm ws}(x)$ is a
sphaleron transition profile function that goes to zero inside the
regions of broken electroweak symmetry and asymptotically to unity
outside, $R$ is a relaxation coefficient for the decay of baryon
number through weak sphaleron transitions, and $n_L(x)$ is the number
density of left-handed doublet fields created by \lq\lq fast" chirality changing processes (see, {\em e.g.}, \cite{Cline:1993bd}). Thus, in order
to obtain nonzero $\rho_B$ inside the bubble of broken electroweak
symmetry, the left-handed density $n_L$ must be non-vanishing
in the plasma at the phase boundary and possibly beyond into the
region of unbroken symmetry. 

In effect, $n_L(x)$ acts as a seed for the $B$-changing weak sphaleron
transitions, and its spacetime profile is determined by the
$CP$-violating sources and the quantum transport of various charges in
the non-equilibrium environment of the plasma. Typical treatments of
these dynamics involve writing down a set of coupled quantum transport
equations (QTEs) for the relevant charges, estimating (or
parameterizing) the relevant transport coefficients, and solving the
system of equations under the appropriate boundary conditions.  

Among the developments in the past decade or so which have made significant impacts on
this program, we identify two that  form the basis of our investigation in this work. First, the authors of Ref.~\cite{Cohen:1994ss} noted
that diffusion of chiral charge ahead of the advancing phase
transition boundary into the region of unbroken symmetry could enhance
impact of baryon number-changing sphaleron processes, thereby leading
to more effective EWB. The second, perhaps less widely-appreciated, development has been the observation by the author of Ref. \cite{Riotto:1998zb}
that the application of equilibrium quantum field theory (QFT) to
transport properties in the plasma is not necessarily appropriate.  In
contrast to equilibrium quantum dynamics, the time evolution of
quantum states during the phase transition is
non-adiabatic. Consequently, scattering processes that drive quantum
transport are no longer Markovian, but rather retain some memory of
the system's quantum evolution. Using the closed time path (CTP)
formulation of non-equilibrium QFT \cite{CTP} to compute the
$CP$-violating source terms in the plasma for the MSSM, the author of
Ref. \cite{Riotto:1998zb} found that these ``memory effects" may lead
to significant resonant enhancements (of order $10^3$) of the sources over
their strength estimated in previous treatments (see, {\em e.g.}, Ref. \cite{Huet:1995sh} and references therein). The authors of Ref. \cite{Carena:2000id,Carena:2002ss} subsequently found that performing an all-orders summation of scattering from Higgs backgrounds reduces  the size of the $CP$-violating sources to some extent, but that the resonant enhancements nonetheless persist. Taken at face value,
these enhancements would imply that successful EWB could occur 
with significantly smaller $CP$-violating phases than previously
believed, thereby evading the present and prospective limits obtained
from EDMs.

To determine whether or not such conclusions are warranted, however,
requires that one treat the other terms in the transport equations in
the same manner as the $CP$-violating sources. Here, we attempt to do so, focusing on the terms that, in previous studies, have governed the relaxation of $n_L(x)$. In particular, chirality-changing Yukawa interactions with the Higgs fields and their spacetime varying vacuum expectation values (vevs) tend to wash out excess $n_L(x)$. In earlier studies---including those in which non-equilibrium QFT has been applied to the $CP$-violating sources---these relaxation terms were estimated using conventional quantum transport theory \cite{Huet:1995sh,Riotto:1998zb,Carena:2000id,Carena:2002ss}. However, if the memory effects that enhance the $CP$-violating sources have a similar effect on these Yukawa terms, then
the net effect on $\rho_B$ may not be as substantial as suggested in
Refs. \cite{Riotto:1998zb,Carena:2000id,Carena:2002ss}.  

The goal of the present study is to
address this question by developing a more comprehensive
treatment of EWB using the CTP formulation of non-equilibrium QFT.  In doing
so, we follow the direction suggested in Ref. \cite{Riotto:1998zb} and
compute the transport coefficients of the chiral charges using the CTP
formalism. To make the calculation more systematic, we identify the relevant energy and time scales that govern finite temperature, non-equilibrium dynamics and develop a power counting in the ratios of small to large scales (generically denoted here as $\epsilon$). As we show below, both the $CP$-violating sources and the driving relaxation terms first arise at ${\cal O}(\epsilon^2)$, and we truncate our analysis at this order. In contrast to the computation of the $CP$-violating sources, the derivation of the relaxation terms requires the use of finite density Green's functions. Given the resulting complexity, we consider here only
the terms in the transport equations that previous authors have
considered the dominant ones, and use our analysis of these terms to
illustrate a method for obtaining a more comprehensive
treatment of the QTEs. To make the phenomenological implications
concrete, we focus on the MSSM, realizing, however, that one may need to
include extensions of the MSSM in order to satisfy the requirements of
a strong first-order phase transition. Finally, we also attempt to identify the
different approximations that have entered previous treatments of EWB,
such as the implicit truncation at a given order in $\epsilon$ and outline additional calculations needed to obtain a comprehensive treatment. 

Based on our  analysis, we find that under that same conditions that lead to resonant enhancements of the $CP$-violating sources, $S^{\CPV}$, one also obtains a similar, resonant enhancement of the driving chirality-changing transport coefficient, ${\bar\Gamma}$. Since $Y_B\sim S^{\CPV}/\sqrt{\bar\Gamma}$, resonant relaxation counteracts the enhanced sources, though some overall enhancement of EWB still persists. Consequently, it will be important in future work to study the other transport coefficients whose impact has been considered sub-leading, since they may be enhanced under other conditions than for the leading terms. From the standpoint of phenomenology, we also illustrate how the implications of EDM searches for EWB depends in a detailed way on the electroweak model of interest as well as results from collider experiments and precision electroweak data. 

In presenting our study, we attempt to be somewhat pedagogical, since the methods are, perhaps, not generally familiar to either the practitioners of field theory or experimentalists.  Most of the formal development appears in Sections \ref{sec:CTP}--\ref{sec:qtes}. In
Section~\ref{sec:CTP} we review the CTP formalism and its application
to the QTEs and discuss in detail the formulation of density-dependent
Green's functions.  In
Section~\ref{sec:source} we compute the $CP$-violating source terms,
providing a check of Ref.~\cite{Riotto:1998zb}, as well as the
transport coefficients of the chiral charge densities. Here, we also
enumerate the approximations used to obtain a set of coupled, linear
differential diffusion equations, discuss their limits of validity,
and identify additional terms (usually assumed to be sub-leading) that we
defer to a future study. In Section~\ref{sec:qtes} we solve these
equations for the baryon density.  A reader primarily interested in the phenomenological implications may want to turn directly to 
Section~\ref{sec:numerics}, which  gives
illustrative numerical studies using the parameters of the
MSSM. A discussion of the implications for EDMs also appears here. Section~\ref{sec:summary} contains a summary and outlook, while several technical details appear in the Appendices.

\section{Non-equilibrium transport: CTP formulation}
\label{sec:CTP}

In what follows, we treat all $CP$-violating and non-topological
chirality-changing interactions perturbatively\footnote{Sphaleron
transitions, however, are manifestly non-perturbative, and we
parameterize their effects in the standard way.}. In contrast to
zero-temperature, equilibrium perturbation theory, however, the
perturbative expansion under non-equilibrium,
$T>0$ conditions requires the use of a more general set of Green's functions
that take into account the non-adiabatic evolution of 
states as well as the presence of degeneracies in the thermal bath. Specifically, the matrix element of any operator 
$\mathcal{O}(x)$ in the interaction representation is given by:
\be
\label{eq:ctp1}
\langle n| S^{\dag}_{\rm int} T\{\mathcal{O}(x) 
S_{\rm int}\} | n\rangle\,  , 
\ee
where 
\be
S_{\rm int} = T\exp\left( i\int d^4x \,{\cal L}_{\rm int}\right)
\ee
for an interaction Lagrangian ${\cal L}_{\rm int}$, $T$ is the
time-ordering operator, and $|n\rangle$ is an in-state. In ordinary,
zero-temperature equilibrium field theory, the assumptions of
adiabaticity and of non-degeneracy of the states $| n\rangle$ implies
that the only impact of $S^{\dag}_{\rm int}$ is the introduction of an
overall phase, allowing one to rewrite (\ref{eq:ctp1}) as:
\be
\label{eq:ctp2}
\frac{\langle n | T\{\mathcal{O}(x)  S_{\rm int}\} | n\rangle}{
\langle n| S_{\rm int} |n\rangle}\,.
\ee
This simplification is no longer valid for non-equilibrium $T>0$ evolution,
and one must take into account the action of $S^{\dag}_{\rm int}$
appearing to the left of $\mathcal{O}(x)$ in (\ref{eq:ctp1}). Doing so is
facilitated by giving every field in $S_{\rm int}$ and $S^{\dag}_{\rm
int}$ a ``$+$" and ``$-$" subscript respectively. The matrix
element in (\ref{eq:ctp1}) then becomes:
\be
\label{eq:ctp3}
\langle n| {\cal P}\left\{ \mathcal{O}(x) 
\exp i\left(\int d^4x\ {\cal L}_{+}-\int d^4x\ 
{\cal L}_{-}\right)\right\}|n\rangle\,,
\ee
where the path ordering operator ${\cal P}$ indicates that all 
``+" fields appear to the right of all ``$-$" fields, with the former
being ordered according to the usual time-ordering prescription and
the latter being anti-time-ordered [here, $\mathcal{O}(x)$ has been taken to
be a ``+" field]. Note that the two integrals in the exponential
in (\ref{eq:ctp3}) can be written as a single integral along a closed
time path running from $-\infty$ to $+\infty$ and then back to
$-\infty$.

Perturbation theory now proceeds from the matrix element
(\ref{eq:ctp3}) along the same lines as in ordinary field theory via
the application of Wick's theorem, but with the more general ${\cal
P}$ operator replacing the $T$ operator. As a result, one now has a
set of four two-point functions, corresponding to the different
combinations of ``+" and ``$-$" fields that arise from
contractions. It is convenient to write them as a matrix ${\widetilde
G}(x,y)$:
\be
\label{eq:ctp4}
\widetilde G(x,y)=
\left(\begin{array}{cc}
G^t(x,y) & -G^<(x,y) \\
G^>(x,y) & -G^{\bar t}(x,y)
\end{array}\right)
\ee
where 
\begin{subequations}
\label{eq:Greens1}
\begin{align}
G^>(x,y) &= \langle \phi_-(x) \phi_+^\dag(y) \rangle \\
G^<(x,y) &= \langle \phi_-^\dag(y) \phi_+(x)\rangle \\
G^t(x,y) &= \langle T\bigl\{\phi_+(x) \phi_+(y)\bigr\}\rangle  = 
\theta(x_0-y_0)G^>(x,y)+ \theta(y_0-x_0)G^<(x,y)\\
G^{\bar t}(x,y) &= 
\langle T\bigl\{\phi_-(x) \phi_-^\dag(y)\bigr\}\rangle  = 
\theta(x_0-y_0)G^<(x,y) + \theta(y_0-x_0)G^>(x,y)\,,
\end{align}
\end{subequations}
 and where the $\langle\ \ \rangle$ denote ensemble averages,
\be
\langle \mathcal{O} (x)\rangle \equiv \frac{1}{Z}
 {\rm Tr}\left[\hat\rho \,  \mathcal{O}(x)\right]\, . 
\ee 
$\hat\rho$ is the density matrix containing 
information about the state of the system.  
In thermal equilibrium  $\hat\rho$ is time-independent and is 
given by  $\hat\rho = e^{-\beta (\op{H} - \mu_i \op{N}_i)}$
for a grand-canonical ensemble. 
Note that the matrix ${\widetilde G}(x,y)$ may be written more compactly as:
 \be
\widetilde G(x,y)_{a b} = 
\langle {\cal P}\left\{\phi_a(x)\phi_b^\dag(y)\right\}\rangle
(\tau_3)_{bb}  \,.
 \ee
The presence of the $\tau_3$ factor is a bookkeeping device to keep
track of the relative minus sign between the ${\cal L}_+$ and ${\cal
L}_-$ terms in Eq. (\ref{eq:ctp3}).
 
The path-ordered two-point functions satisfy the Schwinger-Dyson equations:
\begin{subequations}
\label{eq:sd}
\begin{align}
\label{eq:sda}
{\widetilde G}(x,y) &= {\widetilde G}^0(x,y)+\int d^4w \int d^4 z\  
{\widetilde G}^0(x,w){\widetilde \Sigma}(w,z){\widetilde G}(z,y) \\
\label{eq:sdb}
{\widetilde G}(x,y) &= {\widetilde G}^0(x,y)+\int d^4w \int d^4 z\ 
{\widetilde G}(x,w){\widetilde \Sigma}(w,z){\widetilde G}^0(z,y)\,,
\end{align}
\end{subequations}
where the ``0" superscript indicates a non-interacting Green's
function and where ${\widetilde\Sigma}(x,y)$ is the matrix of interacting
self energies defined analogously to the ${\widetilde G}(x,y)$. An
analogous set of expressions apply for fermion Green's functions, with
an appropriate insertion of $-1$ to account for anticommutation
relations.

\subsection{Quantum Transport Equations from CTP Formalism}
 
The Schwinger-Dyson Eqs.~(\ref{eq:sd}) are the starting
point for obtaining the transport equations governing $n_L(x)$. To do
so, we follow Ref. \cite{Riotto:1998zb} and apply the Klein-Gordon
operator to ${\widetilde G}(x,y)$. Using
\be
\left({{\square}}_x+m^2\right) {\widetilde G}^0(x,y)=  
\left({{\square}}_y+m^2\right) {\widetilde G}^0(x,y)=-i\delta^{(4)}(x-y) 
\ee
gives
\begin{subequations}
\begin{align}
\left({{\square}}_x+m^2\right){\widetilde G}(x,y) &= -i\delta^{(4)}(x-y) -i
\int d^4z\ {\widetilde\Sigma}(x,z){\widetilde G}(z,y) \\
\left({{\square}}_y+m^2\right){\widetilde G}(x,y) &= -i\delta^{(4)}(x-y) -i
\int d^4z\ {\widetilde G}(x,z){\widetilde\Sigma}(z,y)\,.
\end{align}
\end{subequations}
It is useful now to consider the $(a,b)=(1,2)$ components of these equations:
\begin{subequations}
\begin{align}
\label{eq:sdc}
\left({{\square}}_x+m^2\right)G^<(x,y) &= -i\int d^4z\ 
\left[\Sigma^t(x,z) G^<(z,y)-\Sigma^<(x,z) G^{\bar t}(z,y) \right]\\
\label{eq:sdd}
\left({{\square}}_y+m^2\right)G^<(x,y) &= -i\int d^4z\ \left[G^t(x,z) 
\Sigma^<(z,y)-G^<(x,z) \Sigma^{\bar t}(z,y) \right]\,.
\end{align}
\end{subequations}
Subtracting Eq. (\ref{eq:sdd}) from Eq. (\ref{eq:sdc}) and multiplying
through by $i$ gives
\be
\label{eq:sde}
i\left({\square}_x-{\square}_y\right) G^<(x,y)=i\lim_{x\to y} 
\partial_\mu^X\left(\partial^\mu_x-\partial^\mu_y\right)G^<(x,y)\,,
\ee
where $X=(x+y)/2$. However, 
\be
\lim_{x\to y}(\partial_\mu^x-\partial_\mu^y) G^<(x,y) = -i j_\mu(X)\,,
\ee
where $j_\mu(x) = i\langle :\!\phi^\dag(x)\overset{\leftrightarrow}{\partial_\mu}\phi(x)\!:\rangle \equiv(n(x), {\vec j}(x))$, since the ``$+$" and ``$-$" labels
simply indicate the order in which the fields $\phi^\dag(y)$ and
$\phi(x)$ occur and may be dropped at this point. Finally, expressing $G^{t, {\bar t}}(x,y)$ and
$\Sigma^{t,{\bar t}}(x,y)$ in terms of $\theta$-functions as in
Eqs.~(\ref{eq:Greens1}), we obtain from Eq.~(\ref{eq:sde}):
\begin{equation}
\begin{split}
\pd{n}{X_0}+{\mbox{\boldmath$\nabla$}}\mcdot\vect{j}(X) 
= \int d^3 z\int_{-\infty}^{X_0} dz_0\
\Bigl[ \Sigma^>(X,z) G^<(z,X)&-G^>(X,z)\Sigma^<(z,X)\\
+G^<(X,z) \Sigma^>(z,X) &- \Sigma^<(X,z) G^>(z,X)\Bigr]\,.
\label{eq:scalar1}
\end{split}
\end{equation}
Following similar steps, but taking the sum rather than the difference of the components of the Schwinger-Dyson equations involving the $S^{>}(x,y)$ component on the LHS, one obtains the analogous continuity equation 
for Dirac fermions:
\begin{equation}
\begin{split}
\pd{n}{X_0} + {\mbox{\boldmath$\nabla$}}\mcdot\vect{j}(X) =  
-\int d^3 z\int_{-\infty}^{X_0} dz_0\
{\rm Tr}\Bigl[ \Sigma^>(X,z) S^<(z,X)&-S^>(X,z)\Sigma^<(z,X)\\
+S^<(X,z) \Sigma^>(z,X) &- \Sigma^<(X,z) S^>(z,X)\Bigr]\,,
\label{eq:fermion1}
\end{split}
\end{equation}
where
\begin{subequations}
\begin{align}
S^>_{\alpha\beta}(x,y) &=  \langle \psi_{-\alpha}(x) {\bar\psi}_{+\beta}(y)\rangle \\
S^<_{\alpha\beta}(x,y) &= -\langle {\bar\psi}_{-\beta}(y) {\psi}_{+\alpha}(x)\rangle\,,
\end{align}
\end{subequations}
displaying explicitly the spinor indices $\alpha,\beta$.
Note that the overall sign of the RHS of Eqs.~(\ref{eq:scalar1},
\ref{eq:fermion1}) differs from that in Ref. \cite{Riotto:1998zb}
since the definition of our Green's functions $G(x,y)$ and $S(x,y)$
differ by an overall factor of $-i$.

In many extensions of the SM, one encounters both chiral and Majorana fermions, which carry no conserved charge. It is useful, therefore, to derive the analogous continuity equation for the axial current $j_{\mu 5}(x) = \langle {\bar\psi}(x)\gamma_\mu\gamma_5\psi(x)\rangle$. Doing so involves multiplying the Schwinger-Dyson equations by $\gamma_5$, performing the trace, and taking the difference rather than the sum of the components involving $S^{>}(x,y)$ on the LHS. The result is:
\begin{align}
\label{eq:fermion1b}
\pd{n_5}{X_0} + {\mbox{\boldmath$\nabla$}}\mcdot\vect{j}_5(X) = & 2im P(X) \\
&+\int d^3 z\int_{-\infty}^{X_0} dz_0\Tr\Bigl\{\Bigl[\Sigma^>(X,z) S^<(z,X) + S^>(X,z)\Sigma^<(z,X) \nonumber \\
&\qquad\qquad\qquad\qquad-S^<(X,z) \Sigma^>(z,X) - \Sigma^<(X,z) S^>(z,X)\Bigr]\gamma_5\Bigr\}\,, \nonumber
\end{align}
where $P(x) = \langle {\bar\psi}(x)\gamma_5 \psi(x)\rangle$ and $m$ is the fermion mass. In principle, one could evaluate $P(x)$ using path-ordered perturbation theory as outlined above.

\subsection{Power Counting of Physical Scales}

Evaluating the various terms in Eqs.~(\ref{eq:scalar1},
\ref{eq:fermion1}) leads to a system of coupled quantum transport
equations for the charges that ultimately determine $n_L(x)$. On the
LHS of these equations, it is conventional to parameterize $\vect{j} =
- D(\mbox{\boldmath$\nabla$}\!n)$, in terms of the diffusion coefficient
$D$ (whose expressions we take from Ref.~\cite{Joyce:1994zn}).
The RHS involves integrating the products of various Green's functions
and self-energies over the entire history of the system. In practice,
this integral depends on the various time and energy scales associated with non-equilibrium dynamics at finite temperature and density. Here, we observe that there exists a hierarchy among these scales that leads to a natural power counting in their ratios (generically denoted here as
$\epsilon$) and that provides for a systematic expansion of the RHS of the transport equations
(\ref{eq:scalar1}, \ref{eq:fermion1}, \ref{eq:fermion1b}). 

The changing geometry associated with the expanding region of broken symmetry and the spacetime variation of the Higgs vevs leads to a decoherence of states that have, initially, precise energy and momentum. The effect is analogous to the quantum mechanical evolution of a particle in a box of side $L$. If the value of $L$ is changed to $L+\Delta L$ in some time interval $\Delta t$, a state that is initially a stationary state for the original box will become an admixture of the stationary states of new box. The shorter the interval $\Delta t$ or the greater the wavenumber $k$ of the initial state, the smaller the probability will be of finding the particle in the state with the same wavenumber in the new system. The time scale that characterizes this decoherence, $\tau_d$, is naturally given by $\tau_d\sim 1/vk$, where $v=\Delta L/\Delta t$ is the velocity of expansion of the box and $k=p/\hbar$. In the present case, the relevant velocity is just $v_w$, the expanding bubble wall velocity, the relevant wavenumber is $k=\abs{\vect{k}}$. The smaller the velocity or the longer the wavelength, the more adiabatic the dynamics of the expanding bubble become and the longer the decoherence time. Equilibrium dynamics are approached in the adiabatic limit: $\tau_d\to\infty$. The need to employ the CTP formalism follows from being in a situation with $v_w>0$, or $\tau_d <\infty$.

A second time scale that one encounters in quantum transport at the phase boundary arises from the presence of degeneracies among states in the thermal bath that vanish in the $T\to 0$ limit. At finite $T$, for example, a single, on-shell fermion may be degenerate with another  state involving an on-shell fermion-gluon pair---a situation that is forbidden at $T=0$. Interactions of strength $g$ that cause mixing between such degenerate states give rise to thermal---or plasma---widths $\Gamma_p$ of order $\alpha T$ with $\alpha=g^2/4\pi$, and transitions between the degenerate states take place on a plasma time scale $\tau_p$ of order $\sim 1/\Gamma_p$. Again, the use of the CTP formalism is necessitated when $\tau_p <\infty$ or $T> 0$.

A third time scale, which we denote $\tau_{\rm int}$, is associated with the intrinsic frequency $\omega_k$ of the quasiparticle states that characterize the plasma dynamics. This time scale is naturally given by $\tau_{\rm int}\sim1/\omega_k$. In the present case, we note that although the decoherence and plasma times are finite, they are typically  much smaller than $\tau_{\rm int}$. For example, $\tau_{\rm int}/\tau_d = v_w k/\omega_k \leq v_w/c$. Numerical studies indicate that $v_w/c \ll 1$. Similarly, $\tau_{\rm int}/\tau_p = \alpha T/\omega_k$. Since quasiparticle thermal masses are of order $gT$ or larger, one also has that the latter ratio is smaller than unity.  Thus, one is naturally led to expand the RHS of the transport equations in these ratios:
\begin{subequations}
\label{eq:tauratios}
\begin{align}
\label{eq:decratio}
0 &< \tau_{\rm int}/\tau_d \ll 1 \\
\label{eq:plasratio}
0 &< \tau_{\rm int}/\tau_p \ll 1\,.
\end{align}
\end{subequations}

Finally, we observe that the generation of baryon number takes place in an environment of finite, but small particle number (or chiral charge) densities $n_i$ that are associated with chemical potentials $\mu_i$. For the temperatures and densities of interest here, one has $|\mu_i|/T \ll 1$, so that the latter ratio also provides for a natural expansion parameter. Denoting each of the ratios\footnote{For our purposes, it is not necessary to distinguish a hierarchy among the different scale ratios.} in Eq.~(\ref{eq:tauratios}) and $\mu_i/T$ by $\epsilon$, we show below that both the $CP$-violating sources and the relaxation term first arise at ${\cal O}(\epsilon^2)$, and we truncate our analysis at this order. We note that doing so introduces some simplifications into the evaluation of the RHS of the transport equations. For example, both the self energies $\Sigma^\gtrless$ and the Green's functions $G^\gtrless$, {\rm etc.} depend on thermal distribution functions $f(T,\mu_i)$ that differ, in general, from their equilibrium values, $f_0(T,\mu_i)$. The difference $\delta f\equiv f(T,\mu_i)-f_0(T,\mu_i)$ that characterizes the departure from equilibrium will be at least of ${\cal O}(\epsilon)$, since it must vanish in the $v_w\to 0$ limit. We find below that the effect of having $\delta f\not\!= 0$ contributes at higher order in $\epsilon$ than we consider here, so that we may use the equilibrium distribution functions in the Green's functions and self-energies.

\subsection{Green's Functions at Nonzero Temperature and Density}

The computation of the various components of ${\widetilde G}(x,y)$ and
${\widetilde\Sigma}(x,y)$ appearing in Eqs. (\ref{eq:scalar1},
\ref{eq:fermion1}) at nonzero temperature and density requires
knowledge of $(T,\mu_i)$-dependent fermion and boson
propagators. The $T$-dependence of propagators has been studied extensively
(see, {\em e.g.} Ref. \cite{LeBellac} and references therein), while the $\mu_i$-dependence of fermion propagators has been studied in Refs.~\cite{finitemu}. Here we summarize the features of $(T,\mu_i)$-dependent propagators which are important for our subsequent application of the real-time, CTP formalism of Sec.~\ref{sec:CTP}, and give some more technical details in Appendix~\ref{appx:props}. 

For pedagogical purposes, we provide
here a brief derivation of the non-interacting fermion propagator but
only give final results for the case of interacting fermions and
bosons. To do so, we start from the mode
expansions for the field operators appearing in the free Dirac
Lagrangian, $\psi(x)$ and ${\bar\psi}(x)$:
\begin{subequations}
\label{eq:modeexp}
\begin{align}
\psi(x) &=  \int\frac{d^3k}{(2\pi)^3}\frac{1}{2\omega_{\vect{k}}} 
\sum_{\alpha=1,2}\left[b_\alpha(\vect{k}) u(\vect{k},\alpha) 
e^{-ik\cdot x} + d^\dag_\alpha(\vect{k}) 
v(\vect{k},\alpha)e^{ik\cdot x}\right] \\
{\bar\psi(x)} &=  \int\frac{d^3k}{(2\pi)^3}\frac{1}{2\omega_{\vect{k}}} 
\sum_{\alpha=1,2}\left[b_\alpha^\dag(\vect{k}) 
{\bar u}(\vect{k},\alpha)e^{ik\cdot x} + d_\alpha(\vect{k})
{\bar v}(\vect{k},\alpha)e^{-ik\cdot x}\right]\,,
\end{align}
\end{subequations}
where $k^\mu=(\omega_{\vect{k}}, \vect{k})$, $\omega_{\vect{k}}=\sqrt{\abs{\vect{k}}^2 + m^2}$, the mode operators satisfy:
\begin{equation}
\bigl\{b_\alpha(\vect{k}), b_\beta^\dag(\vect{k}^\prime)\bigr\} =  \bigl\{d_\alpha(\vect{k}), d_\beta^\dag(\vect{k}^\prime)\bigr\} = 
(2\pi)^3\delta^{(3)}(\vect{k}-\vect{k}^\prime)2\omega_{\vect{k}}\delta_{\alpha\beta},
\end{equation}
and
\begin{subequations}
\begin{align}
\langle b_\alpha(\vect{k})^\dag b_\beta(\vect{k}^\prime)\rangle &= 
 f(\omega_k,\mu_i) (2\pi)^3 \delta^{(3)}(\vect{k}-\vect{k}^\prime)2\omega_{\vect{k}}\delta_{\alpha\beta}\\
\langle d_\alpha(\vect{k})^\dag d_\beta(\vect{k}^\prime)\rangle &=  
f(\omega_k,-\mu_i) (2\pi)^3 \delta^{(3)}(\vect{k}-\vect{k}^\prime)2\omega_{\vect{k}}
\delta_{\alpha\beta}\ ,
\end{align}
\end{subequations}
with $f(\omega,\mu_i)$ being the non-equilibrium Fermi distribution
function. For our purposes, the relative change $\delta
f(\omega,\mu_i)/f_0(\omega,\mu_i)$ enters the transport equations
multiplying explicit factors of $\Gamma_p$ and either $v_w$ or $\mu$, so that in working to second order in $\epsilon$ we may replace $f$ by the equilibrium
distributions $f_0(\omega,\mu_i)=n_F(\omega-\mu_i)=[e^{(\omega-\mu_i)/T} +
1]^{-1}$. Using the mode expansion (\ref{eq:modeexp}) it is
straightforward to show that 
$S^>(x,y)=\langle \psi(x){\bar\psi}(y)\rangle$ and 
$S^<(x,y)=-\langle \psi(x){\bar\psi}(y)\rangle$  can be expressed 
as ($\lambda$ denotes either ``$>$" or ``$<$"):
\be
\label{eq:slambdafree}
S^\lambda(x,y)=\int {d^4k\over (2\pi)^4} e^{-i{k}\cdot(x-y)}
g_F^\lambda({k}_0,\mu_i)\rho(k_0,\vect{k})\left(\diracslash{k}+m\right)
\,,
\ee
in terms of the free particle spectral density:
\be
 \rho({k}_0, \vect{k}) =  {i\over 2\omega_k}\biggl[
\left({1\over
 {k}_0-\omega_k+i\epsilon}-{1\over {k}_0+\omega_k+i\epsilon}\right)
 -\left({1\over {k}_0-\omega_k-i\epsilon}-
{1\over {k}_0+\omega_k-i\epsilon}\right)\biggr]\  .
\label{eq:spectral1}
\ee
and the functions:
\begin{subequations}
\begin{align}
g_F^>(k_0,\mu_i)&=1-n_F(k_0-\mu_i) \\
g_F^<(k_0,\mu_i)&= -n_F(k_0-\mu_i)\,. 
\end{align}
\end{subequations}
The propagators $S^{t,\bar t}(x,y)$ can now be constructed from the
$S^\lambda(x,y)$ as in Eqs. (\ref{eq:Greens1}).

In the presence of interactions (characterized by a generic coupling
$g$), the fermion propagator becomes considerably more complicated
than given by Eq. (\ref{eq:slambdafree}). In particular, single
fermion states can mix with other multiparticle states in the thermal
bath, leading to the presence of additional poles (the ``hole"
modes) in the fermion propagator \cite{Klimov,Weldon:1989ys}. The general structure of the fermion propagator arising from these effects has been studied extensively at zero density \cite{Weldon:1999th}. In Appendix A we generalize to the
case of non-zero $\mu_i$. For massless fermions, the resulting
propagators are given by:
\be
\label{eq:slambdaint}
S^\lambda(x,y;\mu_i)=\int {d^4k\over (2\pi)^4} e^{-ik\cdot(x-y)} 
g_F^\lambda(k_0, \mu) 
\left[\frac{\gamma_0-\boldgamma\mcdot\vect{\hat k}}{2}\rho_+(k_0, \vect{k}, \mu_i)
+ \frac{\gamma_0+\boldgamma\mcdot\vect{\hat k}}{2}\rho_-(k_0, \vect{k}, \mu_i)\right]
\,,
\ee
where $\vect{\hat k}$ is the unit vector in the $\vect{k}$ direction, and
\begin{equation}
\label{eq:rhoplus}
\begin{split}
\rho_+(k_0,\vect{k},\mu_i) = i\biggl[&\frac{Z_p(k,\mu_i)}{k_0-\mathcal{E}_p(k,\mu_i)}
-\frac{Z_p(k,\mu_i)^*}{k_0-\mathcal{E}_p(k,\mu_i)^*} \\
+ &\frac{Z_h(k,-\mu_i)^*}{k_0+\mathcal{E}_h(k,-\mu_i)^*}
- \frac{Z_h(k,-\mu_i)}{k_0+\mathcal{E}_h(k,-\mu_i)}+F(k_0^*,k,\mu_i)^*-F(k_0,k,\mu_i)\biggr]\,,
\end{split}
\end{equation}
and
\begin{equation}
\label{eq:rhominus}
\rho_-(k_0,\vect{k},\mu_i) = [\rho_+(-k^{0*},\vect{k},-\mu_i)]^*\,.
\end{equation}
%
Here, $\mathcal{E}_p(k,\mu_i)$ and $-\mathcal{E}_h(k,-\mu_i)^*$ are the two (complex) roots (in $k_0$) of the equation:
\be
0 = k_0-k+D_+(k_0, k,\mu_i)+i\epsilon
\ee
where $iD_{\pm}(k_0,k,\mu_i)$ are contributions to the inverse, retarded
propagator proportional to $(\gamma_0\mp\boldgamma\mcdot\vect{\hat k})/2$
arising from interactions.  The function $F(k_0, k,\mu_i)$ gives the
non-pole part of the propagator, and $k=\abs{\vect{k}}$. We find that the resonant contributions to the particle number-changing sources arise from the pole parts of the propagators, so from here on we neglect the terms containing $F(k_0,k,\mu_i)$.

In the limit $g\to 0$, one has $Z_h\to 0$ and $Z_p\to 1$, recovering
the form of the propagator given in Eq.~(\ref{eq:slambdafree}). For
nonzero $g$, however, $Z_h$ is not of order $g^2$ since the particle
and hole modes arise from mixtures of degenerate states. In
particular, at $k=0$ one has $Z_p=Z_h=1/2$. As $k$ becomes large (of
order the thermal mass or larger), $Z_h/Z_p \ll 1$, and the particle
dispersion relation is well-approximated by $\mathcal{E}_p^2 =
\abs{\vect{k}}^2 + m^2(T,\mu_i)$, where $m(T,\mu_i)$ is the thermal mass.
In our particular application to the MSSM, the gaugino $M_i$ masses
will typically be taken to be of order several hundred GeV, and for
the SU(2)$_L\times$U(1)$_Y$ sector, thermal effects do not induce
substantial mass corrections. We find that the gaugino contributions
to the RHS of Eqs. (\ref{eq:scalar1}, \ref{eq:fermion1}) are dominated
by momenta of order $M_i$, so that the hole contributions to the
gaugino $S^{\lambda}(x,y)$ can be neglected. In contrast, for quarks
we find non-negligible contributions from the low-momentum region, so
we retain the full structure given by
Eqs. (\ref{eq:slambdaint}-\ref{eq:rhominus}) in computing their
contributions.

It has been noted in previous studies of quark damping rates that the
one-loop thermal widths $\Gamma_{p,h}= \Imag{\mathcal{E}_{p,h}(k,\mu)}$ are
gauge-dependent (see Ref.~\cite{Braaten:1989mz} and Ref. [3] therein), whereas the thermal masses $m(T,\mu)$ entering
$\mathcal{E}_{p,h}$ are gauge-independent to this order. Gauge-independent
widths can be obtained by performing an appropriate resummation of
hard thermal loops (HTLs) \cite{LeBellac,Braaten:1989mz,Braaten:1991gm}. The latter are associated with momenta
$k_0,k\sim gT$, for which the one-loop functions $D_{\pm}(k_0,k,\mu)$
are of the same order in $g$ as the tree-level inverse propagators. In
what follows, we will estimate the widths $\Gamma_{p,h}$ based on
existing computations~of damping~\cite{Braaten:1992gd, Enqvist:1997ff, Elmfors:1998hh}, deferring a
complete computation of the gauge-invariant, $\mu_i$-dependent
contributions in the MSSM to a future study. In general,
the residues $Z_{p,h}$ also carry a gauge-dependence, and at this time
we are not aware of any HTL resummation that could eliminate this
dependence. In principle, elimination of this gauge-dependence
requires inclusion of one-loop vertex corrections in the computation
of the $\Sigma^\lambda(x,y)$ and $S^\lambda(x,y)$ appearing on the RHS
of Eqs. (\ref{eq:scalar1}, \ref{eq:fermion1}), and
we again defer a complete one-loop computation to a future study.

The derivation of the finite-density scalar propagators proceeds along
similar lines.  Starting from the mode expansion of the free scalar
field $\phi(x)$ in terms of plane-wave solutions to the Klein-Gordon
equation and following analogous arguments as for fermions, one
arrives at the following scalar Green's functions:
\be
G^\lambda(x,y)=\int{d^4k\over (2\pi)^4} e^{-ik\cdot(x-y)} 
g_B^\lambda(k_0, \mu_i)\rho(k_0,\vect{k})
\ee
where the equilibrium distribution functions are:
\begin{subequations}
\begin{align}
g_B^>(\omega, \mu) &=  1+n_B(\omega-\mu_i)\\
g_B^<(\omega, \mu) &= n_B(\omega - \mu_i)\,,
\end{align}
\end{subequations}
with $n_B(x) = 1/(e^{x/T} - 1)$ and $\rho(k_0,\vect{k})$
given by Eq.~(\ref{eq:spectral1}).  As with fermions, one may include
the effect of thermal masses and widths by replacing $m^2\to
m^2(T,\mu_i)$ and $i\epsilon\to i\epsilon+i\Gamma(T,\mu_i)$.

\section{Source Terms for Quantum Transport}
\label{sec:source}

\begin{figure}[!b]
\centering
\begin{picture}(150,180)  
\put(50,50){\makebox(50,50){\epsfig{figure=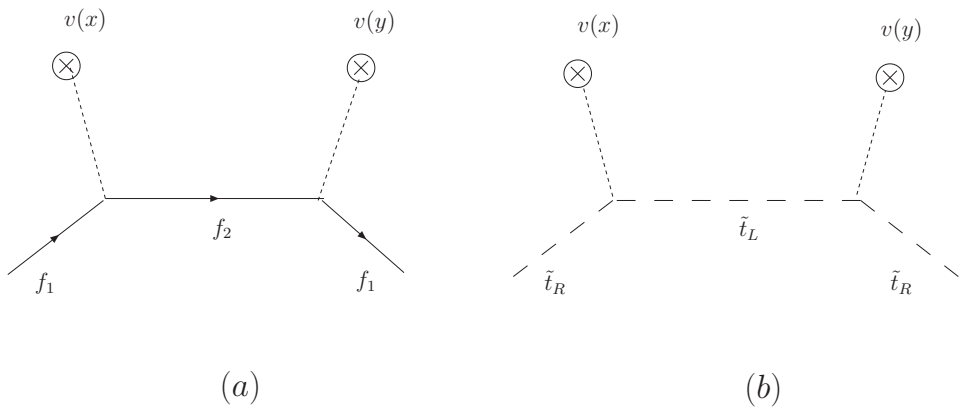,width=13cm}}}
\end{picture}
\caption{
Contributions to the relevant self-energies 
from scattering of particles from the spacetime
varying Higgs vevs. 
\label{fig:graphs1}
}
\end{figure}

The expressions for $G^\lambda(x,y)$ and $S^\lambda(x,y)$ now allow us
to compute the perturbative contributions to the source terms on the
RHS of Eqs. (\ref{eq:scalar1},\ref{eq:fermion1}) starting from a given
electroweak model Lagrangian. Here, we work within the MSSM as an
illustrative case, but emphasize that the methods are general. The
Feynman rules giving the relevant interaction vertices in the MSSM are
taken from Ref.~\cite{Martin:1997ns}, and in what follows, we only write
down those relevant for the computations undertaken here. It is
useful, however, to place our calculation in a broader context by
considering the various classes of graphs that generate different
terms in the QTEs. The simplest topologies are those involving
scattering of particles and their superpartners from the spacetime
varying Higgs vevs (generically denoted $v$) in the plasma
[Fig.~\ref{fig:graphs1}]. 
These graphs give rise to both the $CP$-violating source
terms discussed in Ref. \cite{Riotto:1998zb} as well as terms
proportional to chiral charge. The latter involve the number densities
of at most two different species, such as the left- and right-handed
top quarks [Fig.~\ref{fig:graphs1}(a)] or their superpartners 
[Fig.~\ref{fig:graphs1}(b)]. For purposes of illustration, we follow Ref.~\cite{Riotto:1998zb} and work in a basis of mass eigenstates in the unbroken phase, treating the interactions with the Higgs vevs perturbatively. This approximation should be reasonable near the phase transition boundary, where both the vevs and their rate of change are small, but it clearly breaks down farther inside the bubble wall, where the vevs become large (of order the
phase transition temperature, $T_c$). In general, one would like to perform a resummation to all orders in the vevs, possibly employing the approximation scheme proposed in Refs. \cite{Carena:2000id,Prokopec}. We postpone a treatment of this resummation to a future study\footnote{ The authors of Ref.~\cite{Carena:2000id} find that carrying out such a resummation reduces the resonant enhancements of the $CP$-violating sources, but they did not consider the $CP$-conserving, chirality-changing terms that are our focus here. The consistency of the proposed approximate resummation with our power counting remains to be analyzed.}.

\begin{figure}[!t]
\centering
\begin{picture}(150,200)  
\put(50,90){\makebox(50,50){\epsfig{figure=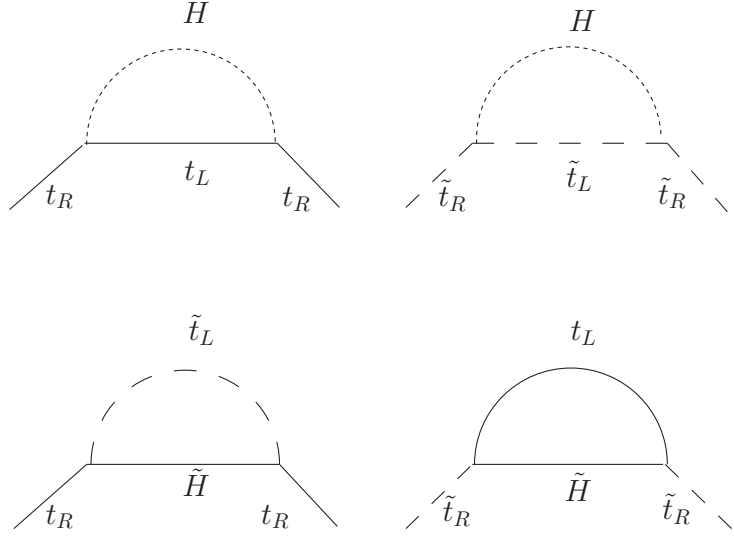,width=10cm}}}
\end{picture}
\caption{
\label{fig:graphs2}
Contributions to the relevant self-energies from Yukawa interactions 
}
\end{figure}

\begin{figure}[!t]
\centering
\begin{picture}(150,230)  
\put(50,90){\makebox(50,50){\epsfig{figure=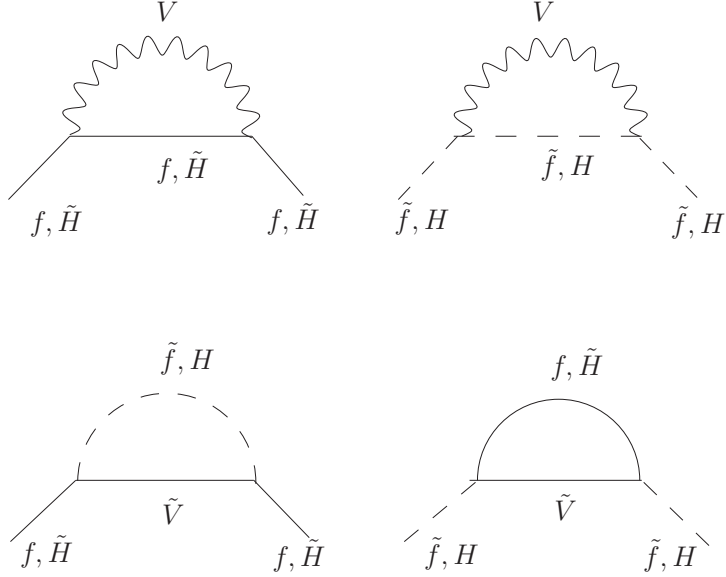,width=10cm}}}
\end{picture}
\caption{Representative contributions to self-energies from
(super)gauge interactions.
\label{fig:graphs3}
}
\end{figure}

Yukawa interactions involving quarks (squarks) and Higgs (Higgsinos)
are illustrated in Fig.~\ref{fig:graphs2} 
(the self-energies $\Sigma^\lambda(x,y)$ are
obtained by amputating the external legs). These interactions cause
transitions such as $f\leftrightarrow f H$, ${\tilde f}\leftrightarrow
{\tilde f} H$, and $f\leftrightarrow {\tilde f}{\tilde
H}$. Contributions from gauge interactions appear 
in Fig.~\ref{fig:graphs3}. The
latter induce transitions of the type $f\leftrightarrow f V$, ${\tilde
f}\leftrightarrow {\tilde f}V$, and $f\leftrightarrow{\tilde f}{\tilde
V}$. 
In general, one expects the Yukawa and gauge
interactions involving three different species to depend on sums and
differences of the corresponding chemical potentials, as in
$\mu_f-\mu_{\tilde f}-\mu_{\tilde V}$ for the supergauge
interactions. In previous studies, it has been assumed that the
gauginos ${\tilde V}$ are sufficiently light and the coefficients of
the corresponding terms in the QTEs sufficiently large than one has
$\mu_{\tilde V}\approx 0$ and $\mu_f\approx \mu_{\tilde f}$. Although
the quantitative validity of this assumption could be explored using
our framework here, we defer that analysis to a future study and take
$\mu_{\tilde V}\approx 0$, $\mu_f\approx \mu_{\tilde
f}$. Consequently, one may, as in Ref. \cite{Huet:1995sh}, define a
common chemical potential for SM particles (including the two Higgs
doublets) and their superpartners.

In previous studies, it has also been assumed -- based largely on simple estimates (see, {\em e.g.}, Ref.~\cite{Huet:1995sh}) -- that the Yukawa interactions of Fig.~\ref{fig:graphs2} are
sufficiently fast that they decouple from the set of QTEs, leading
to relations between the chemical potentials for the Higgs (Higgsino)
fields and those for matter fields. For example, Yukawa interactions
that couple the Higgs doublet fields $H$ with those of the third
generation SU(2)$_L$ doublet quarks, $Q$ with the singlet top quark
supermultiplet field, $T$, generate terms of the form:
\be
\Gamma_Y \left(\mu_Q-\mu_T+\mu_H\right)\,.
\ee
To the extent that $\Gamma_Y$ is much larger than the other transport
coefficients appearing in Eqs. (\ref{eq:scalar1},\ref{eq:fermion1}),
one has $\mu_Q=\mu_T-\mu_H$ plus terms of ${\cal O}(1/\Gamma_Y)$. The
remaining terms in the QTEs will involve the $CP$-violating sources,
sphaleron terms, and terms that couple left- and right-handed chiral
charges, such as $\Gamma_M(\mu_Q-\mu_T)$. Again, this assumption could
be tested using the current framework, but the computation of
$\Gamma_Y$ is considerably more arduous than those discussed below,
where we focus on the $CP$-violating sources and the $\Gamma_M$-type
terms that are generated by the diagrams in Fig.~\ref{fig:graphs1}.

\subsection{Bosons} 

We consider first the scalar interactions in Fig.~\ref{fig:graphs1}(a).
The largest contributions involve the L and R top squarks, ${\tilde
t}_{L,R}$ owing to their large Yukawa coupling, $y_t$. In the basis of weak eigenstates, the relevant interaction
Lagrangian is:
\begin{equation}
\label{eq:scalarhiggs}
\mathcal{L} = y_t\tilde t_L\tilde t_R^*(A_t v_u - \mu^* v_d) + \text{h.c.}\,,
\end{equation}
%
where $v_{u,d}$ are the vevs of $H_{u,d}^0$, and we take $v\equiv\sqrt{v_u^2+v_d^2}$ and $\tan\beta\equiv
v_u/v_d$. Note that in Eq. (\ref{eq:scalarhiggs}) we allow the $v_{u,d}$ to
be spacetime-dependent. In the region of broken electroweak symmetry
and stable vevs, we have $m_t=y_t v_u$.

Using the Feynman rules for path ordered perturbation theory, it is
straightforward to show that the diagrams in Fig.~\ref{fig:graphs1}(a)
generate contributions to ${\widetilde\Sigma}_R(x,y)$ of the form:
\be
\label{eq:sigmascalar}
{\widetilde\Sigma}_R(x,y) = -g(x,y){\widetilde G}_L^0(x,y)
\ee
where
\begin{equation}
\label{eq:gxyscalar}
g(x,y) =  y_t^2\bigl[A_t v_u(x) - \mu^* v_d(x)\bigr] \bigl[A_t^* v_u(y) - \mu v_d(y)\bigr]\,.
\end{equation}
Substituting Eq. (\ref{eq:sigmascalar}) into Eq. (\ref{eq:scalar1}) leads to:
\be
\partial_\mu {\tilde t}^\mu_R(x) = S_{{\tilde t}_R}(x)
\ee
for right-handed top squarks, where ${\tilde t}^\mu_R$ is the
corresponding current density and the source $S_{{\tilde t}_R}(x)$ is
\begin{equation}
\label{eq:scalar2}
\begin{split}
S_{{\tilde t}_R}(x) = -\int d^3z\int_{-\infty}^{x_0}dz_0\ 
\biggl\{[g(x,z)&+g(z,x)]\Real \bigl[G^>_L(x,z)G^<_R(z,x)-G^<_L(x,z)G^>_R(z,x)\bigr] \\
+i[g(x,z)&-g(z,x)]\Imag\bigl[G^>_L(x,z)G^<_R(z,x)-G^<_L(x,z)G^>_R(z,x)\bigr]\biggr\}\,,
\end{split}
\end{equation}
where the L,R subscripts indicate the propagators for the L and R
top squarks.

The first term in the integrand of $S_{{\tilde t}_R}(x)$ is
$CP$-conserving and leads to the $\Gamma_M$-type terms discussed
above, while the second term in the integrand provides the
$CP$-violating sources. We concentrate first on the former. Expanding
$g(x,z)$ about $z=x$ it is straightforward to show that only terms
involving even powers of derivatives survive in $g(x,z)+g(z,x)$. Under
the assumptions of gentle spacetime dependence of the $v_i(x)$ near
the phase boundary, we will neglect terms beyond leading order and
take $g(x,z)+g(z,x)\approx 2 g(x,x)$. Consequently, the
$CP$-conserving source is:
\bea
\label{eq:scalar3}
S_{{\tilde t}_R}^{CP}(x) &\approx& -2g(x,x) \Real
\int d^3z\int_{-\infty}^{x_0}dz_0\  \left [G^>_L(x,z)G^<_R(z,x)
-G^<_L(x,z)G^>_R(z,x)\right]\\
\nonumber
&=& -2g(x,x)\Real\int d^3z \int_{-\infty}^{x_0}dz_0\int{d^4k\over 
(2\pi)^4}\int {d^4q\over (2\pi)^4}
e^{-i(k-q)\cdot(x-z)}\rho_L(k_0,\vect{k})\rho_R(q_0,\vect{q})\\
\nonumber
&&
\ \ \ \times \left[g_B^>(k_0,\mu_L)g_B^<(q_0,\mu_R)-
g_B^<(k_0,\mu_L) g_B^>(q_0,\mu_R)\right]\,,
\eea
with
\be
\label{eq:scalargxx}
g(x,x) = y_t^2 \bigl[|\mu|^2 v_d^2(x) + |A_t|^2 v_u^2(x) -
 2v_d(x) v_u(x) \Real(\mu A_t)\bigr]\,.
\ee
Performing the $d^3z$ integral leads to a $\delta$ function in
momentum space. After carrying out the $d^3q$ integral, 
we perform the contour integrals for $k_0$ and $q_0$, expand to
first order in $\mu_{L,R}/T$, and obtain:
\begin{equation}
\label{eq:scalar5}
\begin{split}
S^{CP}_{\tilde t_R}(x) &= -\frac{1}{T}\frac{N_C y_t^2}{2\pi^2}\abs{A_t v_u(x) - \mu^*v_d(x)}^2 \int_0^\infty\frac{dk\,k^2}{\omega_L\omega_R} \\
&\qquad\times\Imag\biggl\{\frac{\mu_L h_B(\mathcal{E}_L) - \mu_R h_B(\mathcal{E}_R^*)}{\mathcal{E}_L - \mathcal{E}_R^*}  + \frac{\mu_R h_B(\mathcal{E}_R) - \mu_L h_B(\mathcal{E}_L)}{\mathcal{E}_L + \mathcal{E}_R}\biggr\},
\end{split} 
\end{equation}
where
\begin{subequations}
\label{eq:scalardefs}
\begin{align}
\omega_{L,R}^2 &= \abs{\vect{k}}^2 + M_{\tilde t_{L,R}}^2 \\
\mathcal{E}_{L,R} &= \omega_{L,R} - i\Gamma_{L,R} \\
h_B(x) &= -\frac{e^{x/T}}{(e^{x/T} - 1)^2},
\end{align}
\end{subequations}
and $M_{\tilde t_{L,R}},\Gamma_{L,R}$ are the thermal masses and widths 
for the $\tilde t_{L,R}$, and the factor of $N_C$ comes from  
summing over the colors.  Note that, in arriving at Eq. (\ref{eq:scalar5}), we have neglected the $\mu_i$-dependence of the pole residues $Z(T, \mu_{L,R})$, thermal frequencies, $\omega_{ L,R}(T,\mu_{L,R})$, and widths, $\Gamma_{L,R}(T,\mu_{L,R})$. The effect on $S^{CP}_{\tilde t_R}(x)$ of the $\mu_i$-dependence of the residues and thermal frequencies is sub-leading in the gauge and Yukawa couplings, whereas the effect from the thermal widths occurs at leading order. The $\mu_i$-dependence of  $\Gamma_{L,R}(T,\mu_{L,R})$ is simply not known, however, so we do not include it here. A more explicit expression for the dependence of $S^{CP}_{\tilde t_R}(x)$ on the thermal frequencies and widths is given in Eqs. (\ref{appx:scalar5}-\ref{appx:gammascalarpm}) Appendix \ref{appx:expand}.

For purposes of future analysis, it is useful to rewrite
Eq. (\ref{eq:scalar5}) as:
\begin{equation}
\label{eq:scalar6}
S^{CP}_{\tilde t_R} = \Gamma_{\tilde t}^+ (\mu_L + \mu_R) + \Gamma_{\tilde t}^-(\mu_L - \mu_R)\,,
\end{equation}
where
\begin{equation}
\label{eq:gammascalarpm}
\Gamma_{\tilde t}^\pm = -\frac{1}{T}\frac{N_C y_t^2}{4\pi^2}\abs{A_t v_u(x) - \mu^*v_d(x)}^2\!\!\int_0^\infty\!\!\frac{dk\,k^2}{\omega_R\omega_L}\Imag\left\{\frac{h_B(\mathcal{E}_L) \mp h_B(\mathcal{E}_R^*)}{\mathcal{E}_L - \mathcal{E}_R^*} - \frac{h_B(\mathcal{E}_L) \mp h_B(\mathcal{E}_R)}{\mathcal{E}_L + \mathcal{E}_R}\right\}.
\end{equation}

Before proceeding with the $CP$-violating source, we comment briefly
on the structure of Eqs. (\ref{eq:scalar6}-\ref{eq:gammascalarpm}). In
particular, we note that 
\begin{itemize}
\item[(i)] Terms of the type $\Gamma^{+}_{\tilde t}$ are
absent from the conventional QTEs for EWB. It is straightforward to
see that in the absence of interactions that distinguish between
${\tilde t}_L$ and ${\tilde t}_R$, $\Gamma^{+}_{\tilde t}=0$, as the
integrand of Eq.~(\ref{eq:gammascalarpm}) is antisymmetric under
$L\leftrightarrow R$ interchange. In contrast, the transport
coefficient $\Gamma^{-}_{\tilde t}$ is nonzero in the limit of exact
${\tilde t}_L\leftrightarrow {\tilde t}_R$ symmetry. This term
corresponds to the usual damping term in the QTEs associated with
scattering from the Higgs vevs.  
\item[(ii)] In the absence of thermal widths $\Gamma_{L,R}$, the quantity in
brackets in Eq.~(\ref{eq:gammascalarpm}) is purely real, and so the
damping term would be zero. 
\item[(iii)] The structure of the energy
denominators implies a resonant enhancement of the integrand for $
M_{\tilde t_{L}}^2 \sim M_{\tilde t_{R}}^2 $.  A similar effect was observed to occur 
for the $CP$-violating sources (see below) in
Refs.~\cite{Riotto:1998zb,Carena:1997gx}.  The expression in Eq. (\ref{eq:gammascalarpm}) makes it clear that the relaxation terms display a resonant behavior as
well. The resulting quantitative impact of this resonance  baryon asymmetry
is discussed in Sect.~\ref{sec:numerics}.
\end{itemize}
Properties (ii) and (iii) are shared by all source and damping terms,
we discuss  below. Note that the explicit factors of $\mu_{L,R}/T$ and property (ii) imply that, away from the resonance region,
$S^{CP}_{\tilde t_R}$ is $\mathcal{O}(\epsilon^2)$.

The computation of the $CP$-violating source, given by the second term
in Eq.~(\ref{eq:scalar2}), proceeds along similar lines. In this case,
the coefficient $[g(x,z)-g(z,x)]$ vanishes for $x=z$, so we must
retain terms at least to first order in the expansion about $x=z$:
\begin{equation}
\begin{split}
\label{eq:gexpand1}
g(x,z)-g(z,x) & = 2i  y_t^2 \Imag(\mu A_t)\left[v_d(x) v_u(z)
-v_d(z) v_u(x)\right]\\
&= 2i  y_t^2 \Imag(\mu A_t) (z-x)^\lambda\left[ v_d(x)
\partial_\lambda v_u(x)-
v_u(x)\partial_\lambda v_d(x)\right] +\cdots\,,
\end{split}
\end{equation}
where the $+\cdots$ indicate higher order terms in the derivative
expansion that we neglect for the same reasons as discussed
previously. When the linear term in Eq. (\ref{eq:gexpand1}) is
substituted in Eq. (\ref{eq:scalar2}), only the time component yields
a nonzero contribution. The spatial components 
vanish due to the spatial isotropy of the spectral 
density: $g_B^\lambda(k_0,\mu) \rho(k_0, \vect{k})
\equiv g_B^\lambda(k_0,\mu) \rho(k_0, \abs{\vect{k}})$. 
We may then make the replacement:
\begin{equation}
\begin{split}
\label{eq:gexpand2}
g(x,z)-g(z,x) & \rightarrow 2i 
y_t^2\Imag(\mu A_t)\left[v_d(x) {\dot v}_u(x)-{\dot v}_d(x) v_u(x)\right]
\, (z - x)^0 \\
&= 2i y_t^2\Imag(\mu A_t) v(x)^2{\dot\beta}(x) \, (z - x)^0     \ .
\end{split}
\end{equation}
In general, we expect ${\dot\beta}$ to be of order $v_w/c$, so that
the $CP$-violating source is first-order in one of the small expansion
parameters discussed earlier. Consequently, when evaluating this term,
we may neglect the $\mu_{L,R}$-dependence of the
$g_B^\lambda(k_0,\mu)$. After carrying out the ($k_0$,$q_0$) contour
integrals and performing the time integration, we obtain:
\begin{equation}
\label{eq:scalarcp1}
\begin{split}
S_{\tilde t_R}^{\CPV} &= \frac{N_C y_t^2}{2\pi^2}\Imag(\mu A_t)v(x)^2 \dot\beta(x)\int_0^\infty\frac{dk\,k^2}{\omega_R\omega_L}\Imag\biggl\{\frac{n_B(\mathcal{E}_R^*) - n_B(\mathcal{E}_L)}{(\mathcal{E}_L - \mathcal{E}_R^*)^2} + \frac{1+n_B(\mathcal{E}_R) + n_B(\mathcal{E}_L)}{(\mathcal{E}_L + \mathcal{E}_R)^2}\biggr\}.
\end{split}
\end{equation}
Again, property (ii), in conjunction with the factor of ${\dot\beta}\propto v_w$, implies that $S_{\tilde t_R}^{\CPV} $ is $\mathcal{O}(\epsilon^2)$. An expression giving a more explicit dependence on the widths and frequencies appears Eq. (\ref{appx:scalarcp1}) of  Appendix \ref{appx:expand}. 
We note that our result agrees with that of Ref.~\cite{Riotto:1998zb} except for a
different relative sign in front of the $\cos 2\phi$ term of that equation and the overall factor of $N_C$.

\subsection{Massive fermions}

The computations for fermions proceed along similar lines. We consider
first the source terms for Higgsinos. 
We recall that it is useful to redefine the Higgsino fields to remove the
complex phase from the Higgsino mass term:
\be
{\cal L}_{\widetilde H}^{\rm mass} = \mu \left(\psi_{H_d^0}\psi_{H_u^0}-
\psi_{H_d^-}\psi_{H_u^+}\right)
+\mu^\ast \left( 
{\bar\psi}_{H_d^0}{\bar\psi}_{H_u^0}-{\bar\psi}_{H_d^-}{\bar\psi}_{H_u^+}
\right)
\ee
via
\be
\psi_{H_d^{0,-}}\to{\widetilde H}_d^{0,-}\quad\quad 
\psi_{H_u^{0,+}}\to e^{-i\theta_\mu}{\widetilde H}_u^{0,+}
\ee
leading to:
\be
{\cal L}_{\widetilde H}^{\rm mass}=
\abs{\mu}\left({\widetilde H}_d^0{\widetilde H}_u^0-{\widetilde H}_d^-{\widetilde H}_u^+\right)
+ \abs{\mu} \left( {\widetilde H}_d^{0\dag} 
{\widetilde H}_u^{0\dag}-{\widetilde H}_d^{-\dag}{\widetilde H}_u^{+\dag}
\right)\,.
\ee
Defining the four component spinors,
\be
\label{eq:higgsinodef}
\Psi_{\widetilde H^+}  =  \left(\begin{array}{c} 
{\widetilde H}_u^+\\ {\widetilde H}_d^{-\dag}
\end{array}
\right)\quad\quad 
\Psi_{\widetilde H^0}  =  \left(\begin{array}{c} 
-{\widetilde H}_u^0\\ {\widetilde H}_d^{0\dag}
\end{array}
\right)\,
\ee
for the Higgsinos, and 
\be
\label{eq:gauginodef}
\Psi_{\widetilde W^+}= \left(\begin{array}{c} 
{\widetilde W}^+ \\
{\widetilde W}^{-\dag} 
\end{array}
\right)\quad\quad 
\Psi_{\widetilde W^0}=\left(\begin{array}{c} 
{\widetilde W}^3
\\ 
{\widetilde W}^{3\dag}
\end{array}
\right)\quad\quad 
\Psi_{\widetilde B}=\left(\begin{array}{c} 
{\widetilde B}
\\
{\widetilde B}^\dag
\end{array}
\right)\,
\ee
for the gauginos, leads to the Higgsino-gaugino-vev interaction:
\begin{equation}
\begin{split}
\label{eq:higgsinoint1}
{\cal L}^{\rm int} &= 
-g_2\bar\Psi_{\widetilde H^+}\left[v_d(x)P_L+v_u(x) e^{i\theta_\mu}P_R\right]\Psi_{\widetilde W^+}\\
&\quad -\frac{1}{\sqrt{2}}\bar\Psi_{\widetilde H^0}\left[v_d(x)P_L+v_u(x) e^{i\theta_\mu}P_R\right] \left(g_2\Psi_{\widetilde W^0}-
g_1\Psi_{\widetilde B}\right)\ +\ {\rm h.c.}
\end{split}
\end{equation}

Note that the spinors $\Psi_{\widetilde H^0}$ and $\Psi_{\widetilde H^+}$ satisfy a
Dirac equation with Dirac mass  $|\mu|$, even
though the ${\widetilde H}_{d,u}^0$ are Majorana particles. The $\Psi_{\widetilde W^\pm}$ are Dirac particles of mass $M_2$, whereas the $\Psi_{\widetilde W^0}$ and $\Psi_{\widetilde B^0}$ are Majorana particles with Majorana masses $M_2$
and $M_1$, respectively. We also note that the construction of the
Dirac spinor $\Psi_{\widetilde H^0}$ allows one to define a vector charge and
corresponding chemical potential, $\mu_{{\widetilde H}^0}$, for the
neutral Higgsinos, even though they are Majorana particles. In
contrast, there exists no such vector charge for the $\Psi_{\widetilde W^0}$
and $\psi_{\widetilde B^0}$. One may, however, study the quantum transport of the axial charge of the Majorana fermions using Eq. (\ref{eq:fermion1b}). An attempt to do so for the neutral Higgsinos was made in Ref. \cite{Carena:2000id}, though only the $CP$-violating sources were evaluated using non-equilibrium methods. The impact of the corresponding axial charge density on the baryon asymmetry was found to be small. We return to this issue in a future study, and consider only the vector densities below. 

The most straightforward computation is that of the ${\widetilde H}^\pm$
source terms. For notational convenience, we rewrite the chargino
interactions in Eq. (\ref{eq:higgsinoint1}) as:
\be
\label{eq:charginoint}
-g_2 \bar\Psi_{\widetilde H^+}\left[g_L(x)P_L+g_R(x)P_R\right]\Psi_{\widetilde W^+} 
+{\rm h.c.}
\ee
In this case, the self-energy generated by Fig.~\ref{fig:graphs1}(a) is:
\be
{\widetilde \Sigma}_{\widetilde H^\pm}(x,y) =
 -g_2^2 \left[g_L(x)P_L+g_R(x) P_R\right] \,
{\tilde S}_{\widetilde W^\pm} (x,y) \, \left[g_L(y)^*P_R+g_R(y)^* P_L\right]\ \ \ .
\ee
Defining:
\begin{subequations}
\begin{align}
g_A(x,y) & \equiv  \frac{g^2_2}{2}\left[g_L(x) g_L(y)^*+g_R(x) 
g_R(y)^*\right]\\
g_B(x,y) & \equiv \frac{g_2^2}{2}\left[g_L(x) g_R(y)^*+g_R(x) g_L(y)^*\right]\,,
\end{align}
\end{subequations}
we obtain for the RHS of Eq. (\ref{eq:fermion1}):
\bea
\label{eq:fermion2a}
S_{\widetilde H^\pm}(x) & = &  
\int d^3z\int_{-\infty}^0 dz_0 \sum_{j=A,B}\biggl\{ \\
&&
\nonumber
\left[g_j(x,z)+g_j(z,x)\right]\ \ {\rm Re}\ 
{\rm Tr}\ \left[S_{\widetilde W^\pm}^>(x,z) S_{\widetilde H^\pm}^<(z,x)
-S_{\widetilde W^\pm}^<(x,z) S_{\widetilde H^\pm}^>(z,x)\right]_j\\
\nonumber
&+& i \left[g_j(x,z)-g_j(z,x)\right]\ {\rm Im}\  
{\rm Tr}\ \left[S_{\widetilde W^\pm}^>(x,z) S_{\widetilde H^\pm}^<(z,x)
-S_{\widetilde W^\pm}^<(x,z) S_{\widetilde H^\pm}^>(z,x)\right]_j\biggr\}\,,
\eea
where the subscripts \lq\lq A" and \lq\lq B" on the traces denote the
contributions arising from the ${\not\! k}$ and $m$ terms,
respectively, in the spectral function in Eq. (\ref{eq:slambdafree})
(an overall factor of $1/2$ due to the presence of the chiral
projectors $P_{L,R}$ has been absorbed in the definition of the
$g_{A,B}$).

As in the case of the scalar fields, the leading density-dependent,
$CP$-conserving contribution to $S_{\widetilde H^\pm}(x) $ arises from the
term in Eq. (\ref{eq:fermion2a}) containing the $x\leftrightarrow z$
symmetric factors $[g_j(x,z)+g_j(z,x)]$. To lowest order in $v_w$, we
may set $x=z$ in these factors. Using the spectral representation of
the $S^{\lambda}(x,y)$ given in Eq. (\ref{eq:slambdafree}), including
gauge-invariant thermal masses and widths, and expanding to first
order in $\mu_{i}/T$, we obtain the chirality-changing source term:
\be
\label{eq:chargino1}
S^{CP}_{\widetilde H^\pm}(x) = \Gamma_{\widetilde H^\pm}^{+}
\left( \mu_{\widetilde W^\pm}
+
\mu_{\widetilde H^\pm} \right) 
+ 
\Gamma_{\widetilde H^\pm}^{-}
\left( \mu_{\widetilde W^\pm}-\mu_{\widetilde H^\pm}
\right)\,,
\ee
where
\begin{equation}
\label{eq:gammaHiggsinopm}
\begin{split}
\Gamma_{\widetilde H^\pm}^{\pm} = \frac{1}{T}\frac{g_2^2}{2\pi^2}v(x)^2\int_0^\infty\!\frac{dk\,k^2}{\omega_{\widetilde H}\omega_{\widetilde W}} \Imag\biggl\{&\left[\mathcal{E}_{\widetilde W}\mathcal{E}_{\widetilde H}^* - k^2 + M_2\abs{\mu}\cos\theta_\mu\sin 2\beta\right]\frac{h_F(\mathcal{E}_{\widetilde W}) \mp h_F(\mathcal{E}_{\widetilde H}^*)}{\mathcal{E}_{\widetilde W} - \mathcal{E}_{\widetilde H}^*} \\
+ &\left[\mathcal{E}_{\widetilde W}\mathcal{E}_{\widetilde H}
+ k^2 - M_2\abs{\mu}\cos\theta_\mu\sin 2\beta\right]
\frac{h_F(\mathcal{E}_{\widetilde W}) \mp h_F(\mathcal{E}_{\widetilde H})}{\mathcal{E}_{\widetilde W} + \mathcal{E}_{\widetilde H}}\biggr\}\,,
\end{split}
\end{equation}
where the definitions of $\omega_{\widetilde H,\widetilde W}$ and
$\mathcal{E}_{\widetilde H,\widetilde W}$ are analogous to those given
in Eqs. (\ref{eq:scalardefs}) and
\begin{equation}
\label{eq:fermiondefs}
h_F(x) = \frac{e^{x/T}}{(e^{x/T} + 1)^2}\,.
\end{equation}
Also, the factor of $\cos\theta_\mu$ is very nearly 1 for the region of small $\theta_\mu$ in which we find ourselves in subsequent sections. The explicit dependence of $\Gamma_{\widetilde H^\pm}^{\pm}$ on thermal frequencies and widths is given in Eq. (\ref{appx:gammaHiggsinopm}) of Appendix \ref{appx:expand}.

In the present case, we follow Ref. \cite{Huet:1995sh} and assume no
net density of gauginos, thereby setting $\mu_{\widetilde W^\pm}=0$ in
Eq. (\ref{eq:chargino1}) and giving:
\be
\label{eq:chargino2}
S^{CP}_{\widetilde H^\pm}(x) = - \Gamma_{\widetilde H^\pm}\mu_{\widetilde H^\pm}\,,
\ee
with $\Gamma_{\widetilde H^\pm}=\Gamma_{\widetilde H^\pm}^{+}+\Gamma_{\widetilde
H^\pm}^{-}$.  In this case, it is straightforward to obtain the
corresponding source term for the neutral Higgsinos,
\be
\label{eq:neutralino1}
S^{CP}_{\widetilde H^0}(x)=- \Gamma_{\widetilde H^0}\mu_{\widetilde H^0}\,,
\ee
where $\Gamma_{\widetilde H^0}$ can be obtained from the formulae for
$\Gamma_{\widetilde H^\pm}$ by making the following replacements: $g_2\to
g_2/\sqrt{2}$ for ${\widetilde W}^0$ intermediate states and $g_2\to 
g_1/\sqrt{2}$,
$\omw\to\omb$, and $\gamw\to\gamb$ for the ${\widetilde B}$ intermediate
states.

The Higgsino $CP$-violating source arises from the second term in
Eq. (\ref{eq:fermion2a}). As before, we expand the $g_j(x,z)$ to first
order about $x=z$ and observe that only the $x_0-y_0$ component survives
when the $d^3 z$ integration is performed. Also note that $g_A(x,y)-g_A(y,x)=2i\ {\rm
Im} g_A(x,y)=0$ so that only the terms proportional to the Higgsino
and gaugino masses contribute. The result is:
\begin{equation}
\label{eq:chargino3}
\begin{split}
S_{\widetilde H^\pm}^{\CPV}(x) &= \frac{g_2^2}{\pi^2}v(x)^2\dot\beta(x) 
M_2\abs{\mu}\sin\theta_\mu \\
&\qquad\times\int_0^\infty\frac{dk\,k^2}{\omega_{\widetilde H}\omega_{\widetilde W}}\Imag\biggl\{\frac{n_F(\mathcal{E}_{\widetilde W}) - n_F(\mathcal{E}_{\widetilde H}^*)}{(\mathcal{E}_{\widetilde W} - \mathcal{E}_{\widetilde H}^*)^2} + \frac{1-n_F(\mathcal{E}_{\widetilde W}) - n_F(\mathcal{E}_{\widetilde H})}{(\mathcal{E}_{\widetilde W} + \mathcal{E}_{\widetilde H})^2}\biggr\}\,.
\end{split}
\end{equation}
The corresponding expression for
$S_{\widetilde H^0}^{\CPV} (x)$ can be obtained by making the same
replacements as indicated above for the $CP$-conserving terms. The correspondence with the results of Ref. \cite{Riotto:1998zb} can be seen from Eq. (\ref{appx:chargino3}) of Appendix 
\ref{appx:expand}. We again find essential agreement, apart from a sign difference on the $\cos 2\phi$ term.

\subsection{Chiral fermions}

The final source term associated with Fig.~\ref{fig:graphs1}(a)
involves $L$ and $R$ top quarks. At this order, the latter only
contribute a $\mu_i$-dependent $CP$-conserving term. In order to
illustrate the structure of this term that arises when the terms of
${\cal O}(g^2)$ are retained, we employ the interacting fermion
propagators of Eqs. (\ref{eq:slambdaint}-\ref{eq:rhominus}). The
result is:
\be
\label{eq:topmass1}
S^{CP}_{t_R}(x) = \Gamma_{t_R}^{+}\left(\mu_{t_L}+\mu_{t_R}\right)+ 
\Gamma_{t_R}^{-}\left(\mu_{t_L}-\mu_{t_R}\right)\,,
\ee
with
\begin{equation}
\label{eq:gquarkplusminus}
\begin{split}
\Gamma_{t_R}^\pm = \frac{1}{T}\frac{N_C y_t^2 v_u(x)^2}{\pi^2}\int_0^\infty dk\,k^2 \Imag&\biggl\{\frac{Z_p^R(k) Z_p^L(k)}{\mathcal{E}_p^R + \mathcal{E}_p^L}\bigl[h_F(\mathcal{E}_p^L) \mp h_F(\mathcal{E}_p^R)\bigr] \\
&+ \frac{Z_p^L(k) Z_h^R(k)^*}{\mathcal{E}_p^L - \mathcal{E}_h^{R*}}\bigl[h_F(\mathcal{E}_p^L) \mp h_F(\mathcal{E}_h^{R*})\bigr] + (p\leftrightarrow h)\biggr\}\,.
\end{split}
\end{equation}
Here, the ``$p$" and ``$h$" subscripts indicate
contributions from the particle and hole modes, and 
``$L$ " and ``$R$" refer to left- and right-handed quarks. We have not included in our calculation the effects of $\mu_{t_{L,R}}$-dependence of the widths $\Gamma_{p,h}^{L,R}(T,\mu_{t_{L,R}})$, which in principle also enter at this order. For an expanded version of Eq.~(\ref{eq:gquarkplusminus}), including these effects, see Eq. (\ref{appx:gquarkplus}) in Appendix \ref{appx:expand}. 

In the limit of $t_L\leftrightarrow t_R$ symmetry, $\Gamma_{t_R}^+$ vanishes, and $\Gamma_{t_R}^-$ simplifies to:
\begin{equation}
\begin{split}
\Gamma_{t_R}^- = \frac{1}{T}\frac{N_C y_t^2 v_u(x)^2}{\pi^2}\int_0^\infty dk\,k^2\Imag\biggl\{\frac{Z_p(k)^2}{\mathcal{E}_p}&h_F(\mathcal{E}_p) + \frac{Z_h(k)^2}{\mathcal{E}_h}h_F(\mathcal{E}_h) \\
&+ \frac{2Z_p(k) Z_h^*(k)}{\mathcal{E}_p - \mathcal{E}_h^*}\bigl[h_F(\mathcal{E}_p) + h_F(\mathcal{E}_h^*)\bigr]\biggr\}\,.
\end{split}
\end{equation}
We observe that all contributions to the $CP$-violating source terms
and the $\Gamma^{\pm}$ vanish in the limit of zero thermal
widths. Since the widths are generically of order $g^2 T$ (here, $g$
denotes either a gauge or Yukawa coupling), the source terms for the
QTEs are generally fourth order in the couplings. 


\section{Quantum Transport Equations and $\rho_B $}
\label{sec:qtes}

We now discuss diffusion equations for the particle species that
significantly contribute to the density of left-handed doublet fermions
$n_L(x)$ [cf. Eq.~(\ref{eq:rhob1})] that acts as the ``seed'' for
baryogenesis. We subsequently relate $\rho_B$ to $n_L$ and solve
explicitly the equations in the case of a simple geometry and profile
for the bubble wall describing the phase boundary.

\subsection{Solving the diffusion equations}

Using the source terms computed in Section \ref{sec:source}, one can
arrive at a coupled set of differential equations for the various 
particle number densities. These equations simplify considerably under
the assumptions of approximate chemical equilibrium between SM
particles and their superpartners ($\mu_f\approx\mu_{\tilde f}$ with
$\mu_{\tilde V}\approx 0$), as well as the
between different members of left-handed fermion doublets 
($\mu_{W^\pm}\approx 0$). In this case, one obtains transport 
equations for densities associated with different members of a
supermultiplet. This approach is the one followed in
Ref.~\cite{Huet:1995sh}, and for pedagogical purposes we summarize the
development here. 

First, we define the appropriate supermultiplet densities:
\begin{subequations}
\begin{align}
Q &\equiv n_{t_L}+n_{\tilde t_L}+n_{b_L} + n_{\tilde b_L}\\
T &\equiv n_{t_R}+n_{\tilde t_R}\\
B &\equiv n_{b_R}+n_{\tilde b_R} \\
H &\equiv  
n_{H_u^+} + n_{H_u^0} - n_{H_d^-} - n_{H_d^0} +n_{\widetilde H_u^+} - n_{\widetilde H_d^-} +n_{\widetilde H_u^0}- n_{\widetilde H_d^0}\,, 
\end{align}
\end{subequations}
where the Higgsino densities arise from the vector charges $n_{\widetilde H^+}={\bar\Psi_{\widetilde H^+}}\gamma^0\Psi_{\widetilde H^+}$ and
$n_{\widetilde H^0}={\bar\Psi_{\widetilde H^0}}\gamma^0 \Psi_{\widetilde H^0}$ associated with the Dirac fields defined in
Eq. (\ref{eq:higgsinodef}). There are analogous definitions for the
first- and second-generation (s)quarks. Although we do not consider them here, one may also define the corresponding axial charge densities. In the case of the Higgsinos, for example, it will involve the sum, rather than the difference, of the $u$- and $d$-type Higgsino densities\footnote{This density was considered in Ref. \cite{Carena:2000id}, and its overall impact on the baryon asymmetry found to be small.} 

The diffusion equation for a density $n_i$ has the structure:
\be
\partial_\mu J_i^\mu = S_i^{CP} + S_i^{\CPV} + S_i^{\rm sph} \ ,
\ee
where $J_i^\mu$ is the current associated with the density $n_i$, 
$S_i^{CP}$ and $S_i^{\CPV}$ are the source terms computed above, 
and $S_i^{\rm sph}$ is the sphaleron transition term.  
Various derivations of the strong sphaleron term 
appear in the literature, so we do not reproduce them here. However,
we note that the expressions in
Refs. \cite{Giudice:1993bb,Huet:1995sh} have erroneously omitted a
factor of $1/N_C$~\cite{Moore:1997im}.

The $CP$-conserving damping terms $S_i^{CP}$ have been given in
Eqs.~(\ref{eq:scalar6}), (\ref{eq:chargino1}), and
(\ref{eq:topmass1}) to linear order in the appropriate chemical
potentials.  
Assuming local thermal equilibrium we relate the number densities to
the chemical potentials via:
\be
n_i=g_i\ \int_0^\infty {d^3k\over (2\pi)^3} 
\left[N(\omega_k,\mu_i)-N(\omega_k, -\mu_i)\right]\,,
\ee
where $N(\omega,\mu)$ is the appropriate boson or fermion distribution
function and $g_i$ counts the internal degrees of freedom (spin and
color).  Dropping terms of ${\cal O}(\mu_i^3)$, one obtains:
\be
n_i={k_i (m_i/T) \,  T^2\over 6}\mu_i \ , 
\label{eq:muvsn}
\ee
where the factors $k_i (m_i/T)$ are exponentially small in the regime
$m_i/T \gg 1$, and reduce in the massless limit to $k_i(0) = 1$ for
chiral fermions, $k_i (0) =2$ for Dirac fermions, and $k_i (0) =2$ for complex
scalars. In our analysis we keep the full dependence on $m_i/T$:
\begin{equation}
k_i(m_i/T) = k_i(0)\frac{c_{F,B}}{\pi^2}\int_{m/T}^\infty dx\,x\,
\frac{e^x}{(e^x \pm 1)^2}\sqrt{x^2 - m^2/T^2}\,,
\end{equation}
where for fermions (bosons) $c_{F(B)} = 6\,(3)$, and we choose the $+(-)$ sign in the denominator.

Using Eq.~(\ref{eq:muvsn}) in
Eqs.~(\ref{eq:scalar6},\ref{eq:chargino1},\ref{eq:topmass1}), and
defining:
\begin{subequations}
\begin{align}
\Gamma_M^{\pm} &= 
{6\over T^2}\left(\Gamma_{t}^{\pm}+\Gamma_{\tilde t}^{\pm}\right) \\
\Gamma_h &= \frac{6}{T^2} \left( 
\Gamma_{\widetilde H^\pm}+\Gamma_{\widetilde H^0} \right) \,, 
\end{align}
\end{subequations}
the resulting set of coupled transport equations is:
\begin{subequations}
\label{eq:qte}
\begin{align}
\label{eq:qte1a}
\partial^\mu T_\mu &= 
\Gamma_M^{+} \left({T\over k_T}+{Q\over k_Q}\right)-\Gamma_M^{-} 
\left({T\over k_T}-{Q\over k_Q}\right)\\
\nonumber
&\quad -\Gamma_Y\left({T\over k_T}-{H\over k_H}-{Q\over k_Q}\right) +   
 \Gamma_{ss}\left({2Q\over k_Q} -{T\over k_T}+{9(Q+T)\over k_B}\right)+
S_{\tilde t}^{\CPV}\\
\label{eq:qte1b}
\partial^\mu Q_\mu &=  -\Gamma_M^{+} \left({T\over k_T}+
{Q\over k_Q}\right)+\Gamma_M^{-} \left({T\over k_T}-{Q\over k_Q}\right)\\
\nonumber
&\quad +\Gamma_Y\left({T\over k_T}-{H\over k_H}-{Q\over k_Q}\right)-
2\Gamma_{ss}\left({2Q\over k_Q} 
-{T\over k_T}+{9(Q+T)\over k_B}\right)-
S_{\tilde t}^{\CPV}\\
\label{eq:qte1c}
\partial^\mu H_\mu &=  -\Gamma_h{H\over k_H}-\Gamma_Y\left({Q\over k_Q}+{
H\over k_H}-{T\over k_T}\right)+S_{{\widetilde H}}^{\CPV}\,,
\end{align}
\end{subequations}
where $\Gamma_{ss}=6 \kappa' \frac{8}{3} \alpha_{s}^4 T$, with $\kappa' \sim 
O(1)$. 

In writing down Eqs. (\ref{eq:qte1a}-\ref{eq:qte1c}), we have also
included the $H {\tilde q}_L {\tilde q}_R$ and $H q_L q_R$ Yukawa
interaction term that arises from Fig. 2, though we have not computed
the corresponding transport coefficient $\Gamma_Y$. 
In this work we will again follow the authors of
Ref. \cite{Huet:1995sh}, who estimate $\Gamma_Y \gg \Gamma_M^{-}$.
For $\kappa' \sim {\cal O}(1)$, one also has $\Gamma_{ss} \gg
\Gamma_M^{\pm}$.  These facts allow one to algebraically relate the
densities $Q$ and $T$ to $H$, by 
setting the linear combinations multiplying $\Gamma_Y$ and
$\Gamma_{ss}$ equal to  $\delta_Y=O(1/\Gamma_Y)$ and 
$\delta_{ss} = O(1/\Gamma_{ss})$,
respectively. 
One then obtains:
\begin{subequations}
\label{eq:qte2}
\begin{align}
\label{eq:qte2a}
Q & =   {(k_B-9k_T)k_Q\over (9k_T+9k_Q+k_B)k_H} \,  H 
+ \alpha_{QY}\delta_Y+\alpha_{Qs}\delta_{ss}\\
\label{eq:qte2b}
T & =   {(9k_T+2k_B)k_T\over (9k_T+9k_Q+k_B)k_H} \, H + 
\alpha_{TY}\delta_Y+\alpha_{Ts}\delta_{ss}\,,
\end{align}
\end{subequations}
with known coefficients $\alpha_{QY,Qs,TY,Ts}$.
Taking 2 $\times$ [Eq. (\ref{eq:qte1a})] $+$ [Eq. (\ref{eq:qte1b})]
$+$ [Eq. (\ref{eq:qte1c})],
introducing the
diffusion approximations $\vect{T}=-D_q\mbox{\boldmath$\nabla$}T$,
$\vect{Q}=-D_q\mbox{\boldmath$\nabla$} Q$,
$\vect{H}=-D_h\mbox{\boldmath$\nabla$} H$, and using
Eq.~(\ref{eq:qte2}) leads to:
\be
\label{eq:qte4}
{\dot H} -{\bar D} \nabla^2 H +{\bar \Gamma} H-{\bar S}= 
\mathcal{O}(\delta_{ss}, \delta_Y) 
\ \ \ ,
\ee
where\footnote{Our expressions differ from those 
 in Ref.~\cite{Huet:1995sh}, which we believe result from an algebraic error. The numerical impact of this difference, however, is not significant.}
\begin{subequations}
\begin{align}
\label{eq:qte5}
{\bar D}& =  {(9k_Q k_T + k_B k_Q + 4k_T k_B)D_q +k_H(9k_T+9k_Q+k_B)D_h\over
9k_Qk_T + k_Bk_Q + 4k_T k_B +k_H(9k_T+9k_Q+k_B)}\\
{\bar \Gamma} & =  \frac{(9k_Q+9k_T+k_B)(\Gamma_M^{-}+\Gamma_h) - (3k_B+9k_Q-9k_T)\Gamma_M^{+}}{9k_Qk_T + k_Bk_Q + 4k_T k_B +k_H(9k_T+9k_Q+k_B)} \\
{\bar S} & =  {k_H (9k_Q+9k_T+k_B)\over 
9k_Qk_T + k_Bk_Q + 4k_T k_B +k_H(9k_T+9k_Q+k_B)}
\left(S_{\tilde t}^{\CPV}+S_{\widetilde H}^{\CPV}\right)\,.
\end{align}
\end{subequations}
The subleading terms $\delta_{Y,ss}$ can be determined by use of
Eqs.~(\ref{eq:qte2}) in
Eqs.~(\ref{eq:qte1a},\ref{eq:qte1b}). We include the effect of
$\delta_{ss}$ in our final expression for $\rho_B$~\cite{Huet:1995sh},
although its  effect is negligible in the relevant MSSM
parameter region.

Equation (\ref{eq:qte4}) can now be solved for a given set of
assumptions about the geometry of the bubble wall.  Again, for
clarity of illustration, 
we will work in a framework that allows us to carry analytic calculations as far as possible,
leaving to the future a numerical solution of the
equations for a realistic wall geometry and profile.  First, as commonly done in earlier
studies, we 
ignore the wall curvature, thereby reducing the problem to a one-dimensional one in
which all relevant functions depend on the variable $\bar{z} =
|\vect{x} + \vect{v}_{w} t|$, where $\vect{v}_{w}$ is the wall velocity.
Thus,  $\bar{z} < 0$ is associated with the unbroken phase, $\bar{z} > 0$
with the broken phase, and the boundary wall extends over $0 < \bar{z}
< L_{w}$.   Second, we take  the relaxation term $\bar{\Gamma}$ to be nonzero and
constant for $\bar{z} > 0$. The resulting solution for $H$ in the unbroken phase   
$\bar{z} < 0$ (related to $\rho_B$ as shown below) is:
\be
\label{eq:sol1}
H(\bar{z}) = {\cal A} \,  e^{ v_{w}\bar{z}/\bar{D}} \,
\ee
with
\be
\label{eq:sol2}
{\cal A} = 
\frac{1}{\bar{D} \kappa_{+}} \, 
\int_{0}^{\infty} \ \bar{S}(y) \,  e^{- \kappa_+  y} \
d y \qquad \qquad 
\kappa_+ = \frac{v_w + \sqrt{v_w^2 + 4 \bar{\Gamma} \bar{D}}}{2 \bar{D}}
\simeq 
\sqrt{\frac{\bar{\Gamma}}{\bar{D}}}\ .
\ee
The above equation is valid for any shape of the source
$\bar{S}(\bar{z})$. For simplicity, however,  we 
assume a simple step-function type behavior  for the
source: $\bar S$ nonzero and constant for 
$0 < \bar{z} < L_{w}$.
Specializing to this case of constant sources
in $0 < \bar{z} < L_{w}$, using $4 \bar{D} \bar{\Gamma} \gg v_w^2$,
$L_w \sqrt{\bar{\Gamma}/\bar{D}} \ll 1$, and taking 
$\bar{\Gamma} = r_{\Gamma} \, (\Gamma_h + \Gamma_{M}^{-})$ from
Eq.~(\ref{eq:qte5}), we arrive at:
\be
\label{eq:sol3}
{\cal A} = k_H \, L_w \, \displaystyle\sqrt{\frac{r_\Gamma}{\bar{D}}} \  
\frac{ S_{\tilde{H}}^{\CPV} + S_{\tilde{t}}^{\CPV}  
}{ 
\sqrt{\Gamma_h + \Gamma_{M}^{-}}} \ .   
\ee
When evaluating the source terms $S_{\tilde{H}}^{\CPV},
S_{\tilde{t}}^{\CPV}$ [see
Eqs.~(\ref{eq:scalarcp1}),(\ref{eq:chargino3})] for this simple
profile one has to use $\dot{\beta} = v_{w} \Delta \beta/L_w$:
thus ${\cal A}$ is explicitly proportional to $v_w$ and is only
weakly dependent on $L_{w}$.  Solutions for $Q$ and $T$ are then
obtained via Eqs.~(\ref{eq:sol1}) and (\ref{eq:qte2}).

\subsection{The baryon density $\rho_{B}$}

Neglecting the wall curvature and assuming a step-function 
profile for the weak sphaleron rate, the baryon density satisfies 
the equation~\cite{Cline:2000nw,Carena:2002ss}:
\be
\label{eq:rhob2}
D_q \rho_B '' (\bar{z}) - v_{w} \rho_B ' (\bar{z}) 
- \theta(-\bar{z}) \, {\cal R} \, \rho_B 
= \theta(-\bar{z})
\, \frac{n_F}{2} \Gamma_{\rm ws} n_L(\bar{z}) 
\ , 
\ee
where $n_F$ is the number of fermion families and the relaxation term 
is given by~\cite{Cline:2000nw}:
\be
{\cal R} = \Gamma_{\rm ws}\, \left[
\frac{9}{4} \, \left(1 + \frac{ n_{\rm sq}}{6}\right)^{-1} + \frac{3}{2}
\right] \ , 
\ee
where $n_{\rm sq}$ indicates the number of light squark flavors, and
the weak sphaleron rate is given by $\Gamma_{\rm ws} = 6 \kappa
\alpha_w^5 T$, with $\kappa \simeq 20$~\cite{wsrate}. 

The solution to Eq.~(\ref{eq:rhob2}) in the broken phase, eventually
growing into the Universe, is constant and given by:
\be
\label{eq:rhob3}
\rho_B = - \frac{n_F \Gamma_{\rm ws}}{2 v_{w}} \, \int_{-\infty}^{0}  \ 
n_L (x) 
\,  e^{x \, {\cal R}/v_w} \, 
dx  \ . 
\ee
Neglecting leptonic contributions, $n_L$is given in the unbroken phase 
by the sum of left-handed quark densities over the three
generations ($Q_{1L}, Q_{2L}, Q$).  Since appreciable densities of
first and second generation quarks are only generated via strong
sphaleron processes, it is possible to express $Q_{1L}$ and $Q_{2L}$
in terms of $Q$ and $T$, in such a way that $n_L =
Q + Q_{1L} + Q_{2L} = 5 Q + 4 T$~\cite{Huet:1995sh}.  
Using then Eq.~(\ref{eq:qte2}) one obtains :
\begin{equation}
\label{eq:rhob4a}
n_L =  -H \, \left[ r_1 + r_2 \, \frac{v_w^2}{\Gamma_{\rm ss} \, \bar{D}} 
\left(1- \frac{D_q}{\bar{D}} \right) \right]\,,
\end{equation}
where
\begin{subequations}
\begin{align}
\label{eq:rhob4b}
r_1 &= \frac{9 k_Q k_T - 5 k_Q k_B - 8  k_T k_B}{k_H (9k_Q+9k_T+k_B)} \\
\label{eq:rhob4c}
r_2 &=  
\frac{k_B^2 (5k_Q+4k_T)(k_Q + 2 k_T)}{k_H (9k_Q+9k_T+k_B)^2} \ ,  
\end{align}
\end{subequations}
and finally, in the broken phase:
\begin{equation}
\label{eq:rhob5}
\begin{split}
\rho_B (\bar{z} > 0) &= \frac{n_F}{2} {\cal A}  \
\left[ r_1  \Gamma_{\rm ws}
+ r_2 \, \frac{\Gamma_{\rm ws}}{\Gamma_{\rm ss}}\frac{v_w^2}{\bar D} 
\left(1- \frac{D_q}{\bar{D}} \right) \right] \, \frac{2\bar D}{v_w\Bigl[v_w + \sqrt{v_w^2 + 4\mathcal{R}D_q}\Bigr] + 2\mathcal{R}\bar D} \\
&=\frac{n_F}{2} {\cal A}  \
\left[ r_1  \Gamma_{\rm ws}
+ r_2 \, \frac{\Gamma_{\rm ws}}{\Gamma_{\rm ss}}\frac{v_w^2}{\bar D} 
\left(1- \frac{D_q}{\bar{D}} \right) \right] \, 
\frac{\bar D}{v_w^2  + {\cal R}(\bar{D} + D_q)} \,, 
\end{split}
\end{equation}
where the second line is true in the limit $v_w^2\gg 4 D_q\mathcal{R}$, which holds for the parameters we have chosen in this calculation. The contribution from the first term in Eq.~(\ref{eq:rhob5}) is linear in $v_w$, due to the 
linear dependence on $v_w$ contained in the ${\dot\beta}$ appearing in the $CP$-violating sources. 
The second term is suppressed by two additional powers of $v_w$ and generally leads to a negligible contribution to $\rho_B$ in the MSSM case (see discussion below). It
could, however, be dominant in the case of heavy degenerate $\tilde{t}_L$ and
$\tilde{t}_R$, which leads to $r_1\sim 0$ ~\cite{Huet:1995sh}. 

The central feature emerging from the above discussion is that the net
baryon density is proportional to ${\cal A} \sim S^{\CPV}/\sqrt{\Gamma}$.  A large relaxation rate $\Gamma$ for the relevant
charges will suppress the overall baryon asymmetry.  While in
Refs.~\cite{Riotto:1998zb,Carena:1997gx} it was pointed out how a
non-equilibrium quantum transport  could result in a resonant
enhancement of $S^{\CPV}$, we observe here that similar resonance effects in the relaxation terms will mitigate the impact of the enhanced sources.
In the next section we discuss the numerical impact within the MSSM, but caution that reaching definitive conclusions will require computing the other transport coefficients, such as $\Gamma_Y$, within the same framework.

\section{Baryogenesis and Electroweak Phenomenology within the MSSM}
\label{sec:numerics}

The results derived in the previous Sections allow us to perform an illustrative, 
preliminary analysis of baryogenesis within the MSSM.  This should be
taken as an exploration, whose robustness will be tested once we
implement the next steps in our treatment of the source terms in the
transport equations.  With this caveat in mind, we  explore the
connections between electroweak  baryogenesis and  phenomenology  
within the MSSM, focusing in particular on the implications for EDM searches.
Throughout, we assume that all the terms in the Higgs scalar potential
and all gaugino masses are real, while all the $A$-parameters (trilinear
scalar couplings) are equal at the GUT scale, therefore sharing the
same phase $\phi_A$.  In this case,  the baryon asymmetry and EDMs are
sensitive to the two independent $CP$ violating phases $\phi_\mu$ and
$\phi_A$.

From the structure of Eqs.~(\ref{eq:rhob5},\ref{eq:sol3}) and
(\ref{eq:scalarcp1},\ref{eq:chargino3}) we can write the 
baryon-to-entropy density ratio\footnote{
We evaluate the entropy density at the electroweak phase 
transitions via $s = (2 \pi^2)/45 \times  g_{\rm eff} (T) \,  T^3$, 
with $g_{\rm eff} = 130.75$, resulting in $s = 57.35 \, T^3$. 
Similarly, to convert the present ratio $\rho_B/n_\gamma$
to $Y_B$,  we  use the relation $s = 7.04 \, n_\gamma$. 
}
$Y_{B} \equiv \rho_B/s$ as:
\be
\label{eq:pheno1} 
Y_{B} = F_1\,  \sin \phi_\mu \ + \  F_2\, \sin \left( \phi_\mu +
\phi_A \right)  \,, 
\ee
where we have isolated the dependence on the phases $\phi_\mu$ and
$\phi_A$.  The first term that contains $F_{1}$ stems from the Higgsino source, while the $F_{2}$ term arises from the squark source.

The functions $F_{1}$ and $F_{2}$ display a common overall dependence on bubble
wall parameters ($v_w$, $L_w$, $\Delta \beta$), while having distinct
dependence on other MSSM mass paramters such as $|\mu|$, the soft mass parameters
for gauginos ($M_{1,2}$) and squarks
($M_{\tilde{t}_L},M_{\tilde{t}_R}$), the triscalar coupling
$|A_t|$, and $\tan \beta$.  In order to assess the size of $Y_B$ and
the impact on $CP$-violating phases, we must choose a reference region
in the MSSM parameter space, and we follow two obvious guidelines: (i)
we require that $v(T_c)/T_c \gsim 1$, so that the baryon asymmetry is
not washed out in the broken symmetry phase; (ii) we require no
conflict with precision electroweak physics and direct collider
searches. Both criteria lead to non-trivial restrictions.

\begin{figure}[!t]
\centering
\begin{picture}(300,170)  
\put(0,60){\makebox(50,50){\epsfig{figure=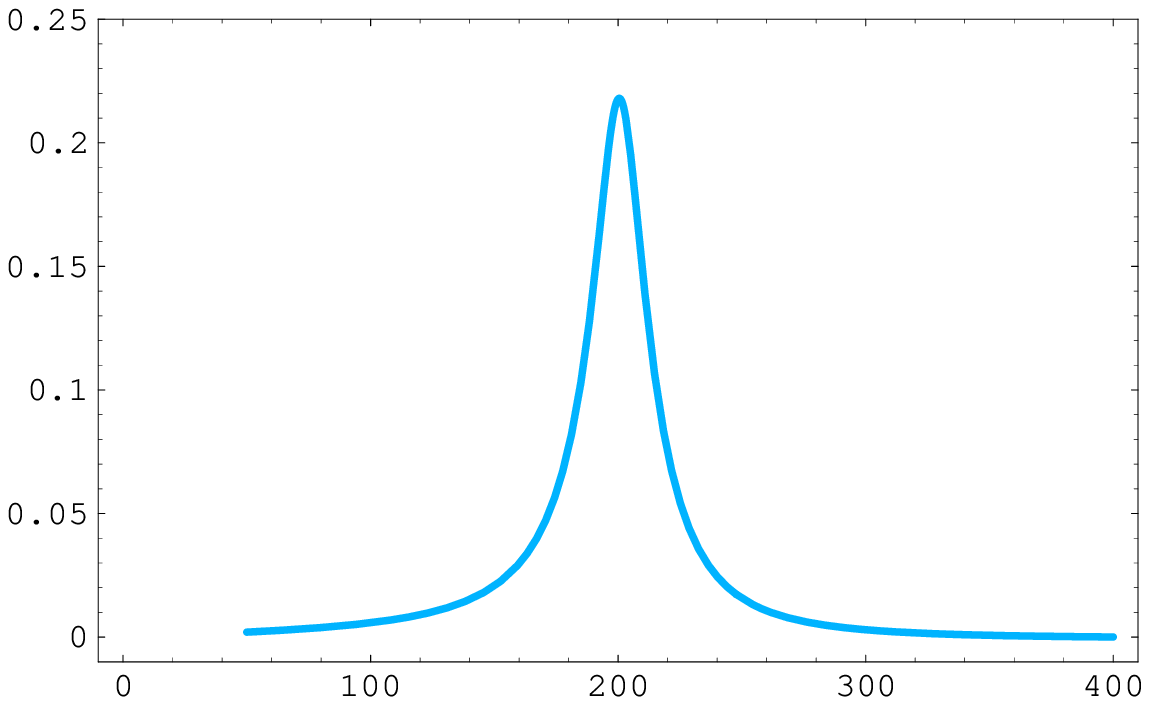,width=7.5cm}}}
\put(110,10){{\scriptsize $\displaystyle |\mu| \ ({\rm GeV})$ }}
\put(350,10){{\scriptsize $\displaystyle |\mu| \ ({\rm GeV})$ }}
\put(240,60){\makebox(50,50){\epsfig{figure=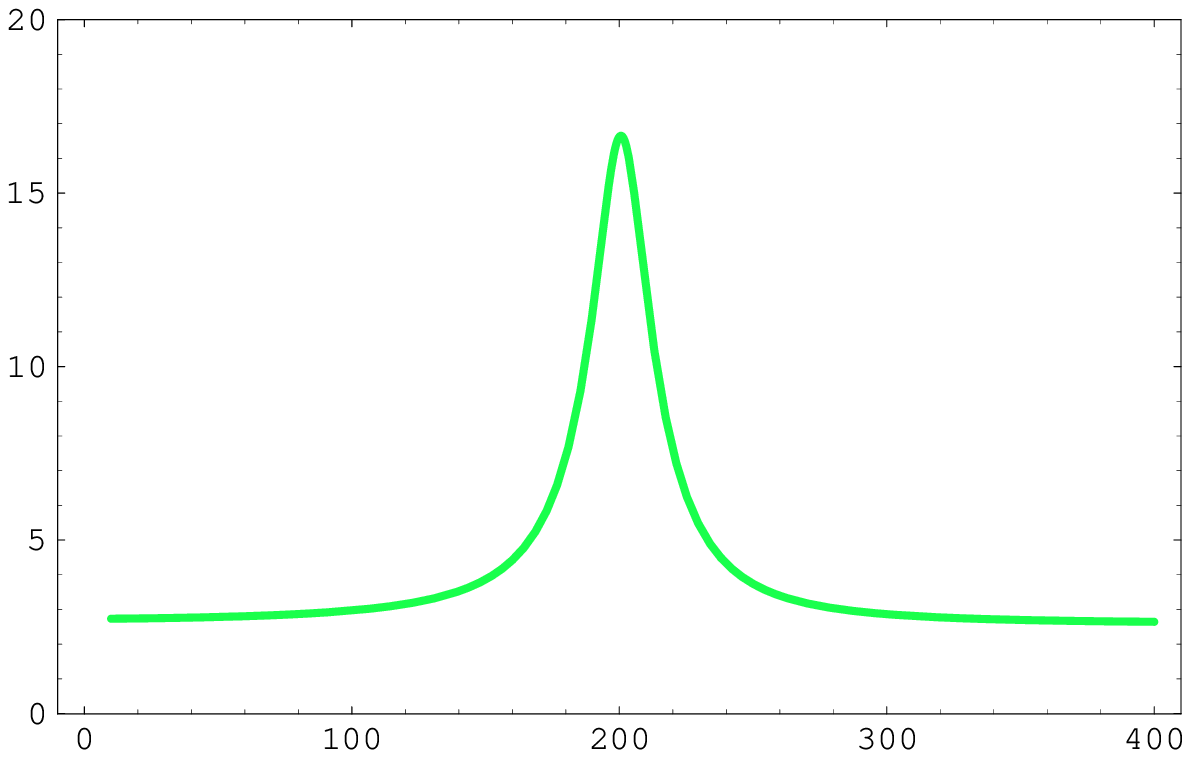,width=7.5cm}}}
\put(-90,160){{\small  $ \hat{S}_{\tilde{H}} $}}
\put(150,160){{\small $ R_{\Gamma} $ }}
\end{picture}
\caption{
Left panel: $CP$-violating Higgsino source
$ \hat{S}_{\widetilde{H}} = 
-S_{\widetilde{H}}^{CP\!\!\!\!\!\!\!\!\mbox{\normalsize$\diagup$}}/(v^2 \dot{\beta} \sin \phi_\mu) $, 
as a function of $|\mu|$.  
Right panel: relaxation rate
$ R_{\Gamma} = (\Gamma_h + \Gamma_{M}^{-})/(\Gamma_h + \Gamma_m)_{H.N.}$, 
normalized to the value used in~\cite{Huet:1995sh},
as a function of $|\mu|$.  We have taken $M_2 = 200\text{ GeV}$, and the values of all other parameters as indicated in the text.
\label{fig:sg}
}
\end{figure}

The condition of a strongly first-order phase transition [$v(T_c)/T_c
\gsim 1$] requires light scalar degrees of freedom coupling to the
Higgs sector.  It has been shown \cite{Carena:1997ki,Laine:1998qk}
that within the MSSM the only viable candidate is a light top quark,
which should be mainly right-handed ($\tilde{t}_R$) in order to avoid
large contributions to the $\rho$ parameter.  Quantitatively, for
lightest Higgs boson mass $m_h \lsim 120$ GeV, one needs $100 \
\mbox{GeV} \lsim m_{\tilde{t}} < m_t$, and sufficiently small stop
mixing parameter $|A_t - \mu/\tan \beta | \lsim 0.6
M_{\tilde{t}_L}$~\cite{Carena:1997ki}.  Moreover, present experimental
limits on $m_h$ and the constraint $v(T_c)/T_c \gsim 1$ jointly
require either values of $\tan \beta >5$ or $M_{\tilde{t}_L}$ in the
multi-TeV region~\cite{Carena:2000id}.  Based on these considerations,
for illustrative purposes we work with the following values of MSSM
parameters at the electroweak scale: $M_{\tilde{t}_R} = 0$,
$M_{\tilde{t}_L} = 1$ TeV, $|A_t| = 200$ GeV, $M_{2} = 200$ GeV, $\tan
\beta = 10$.  We also take for the $CP$-odd Higgs mass $m_A = 150$ GeV,
which translates into $\Delta \beta \sim 0.015$~\cite{bubble}.   
We vary in the plots the scale $|\mu|$, in order to display the
resonant behavior for $|\mu| \sim M_2$.  Finally, for the bubble wall
parameters we adopt the central values $v_w = 0.05$ and $L_w = 25/T$
~\cite{bubble}.

\begin{figure}[!t]
\centering
\begin{picture}(300,170)  
\put(0,60){\makebox(50,50){\epsfig{figure=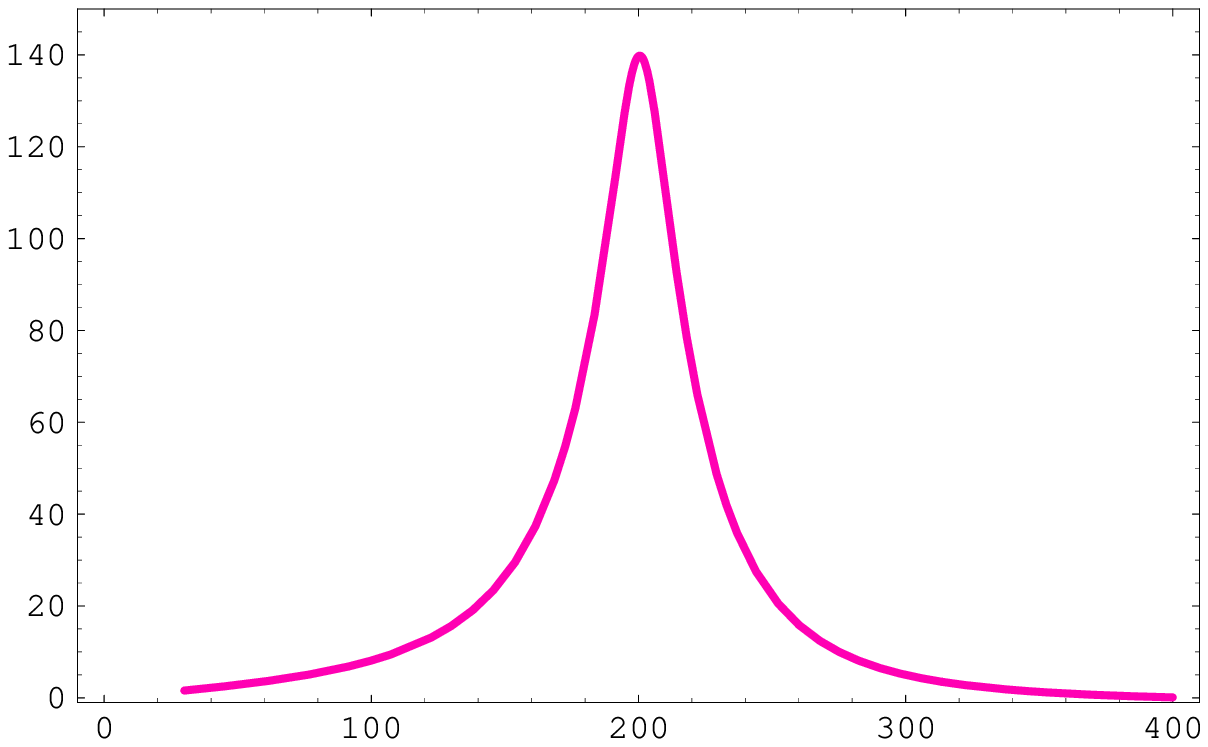,width=7.5cm}}}
\put(240,60){\makebox(50,50){\epsfig{figure=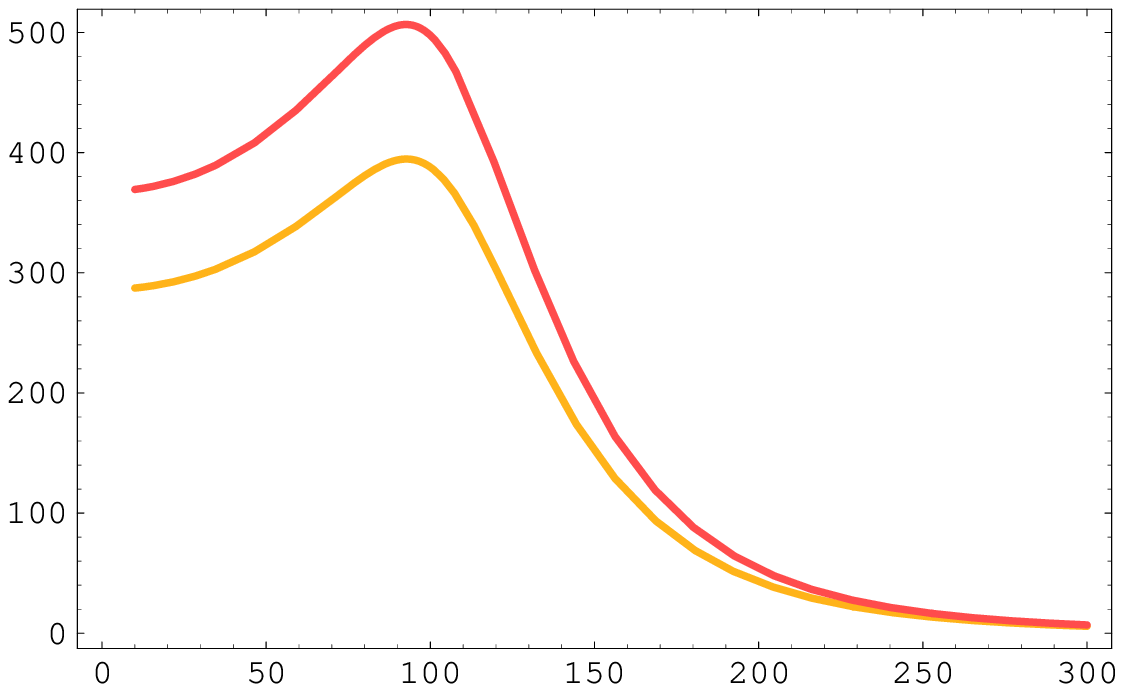,width=7.5cm}}}
\put(110,10){{\scriptsize $\displaystyle |\mu| \ ({\rm GeV})$ }}
\put(350,10){{\scriptsize $\displaystyle M_{\tilde{t}_L} \ ({\rm GeV})$ }}
\put(-100,160){{\scriptsize 
$\displaystyle\frac{F_1}{Y_{B}^{\rm WMAP}}$ 
}}
\put(145,160){{\scriptsize 
$\displaystyle\frac{F_2}{Y_{B}^{\rm WMAP}}$ 
}}
\end{picture}
\caption{Left panel: Higgsino contribution to $Y_B$ 
({\it Cf}  Eq.~(\ref{eq:pheno1})), normalized to the observed value.  $F_{1}$
displays residual resonant behavior for $|\mu| \sim M_2$. All other
input parameters are given in the text.  Right panel: Stop
contribution to $Y_B$ ({\it Cf}  Eq.~(\ref{eq:pheno1})) normalized to
the observed value.  The upper curve is for 
$M_{\tilde{b}_L}=M_{\tilde{t}_L}$, while the lower one is for
$M_{\tilde{b}_L} \gg M_{\tilde{t}_L}$.  We have taken here
$M_{\tilde{t}_R}= 100$ GeV, $| \mu | = 200$ GeV, and have allowed
$M_{\tilde{t}_L}$ to reach unrealistically low values to explore the
size of the squark resonance.  For realistic input parameters $F_2 \ll
F_1$.
\label{fig:F1}
}
\end{figure}

\begin{figure}[!t]
\centering
\begin{picture}(300,180)
\put(120,68){\makebox(50,50){\epsfig{figure=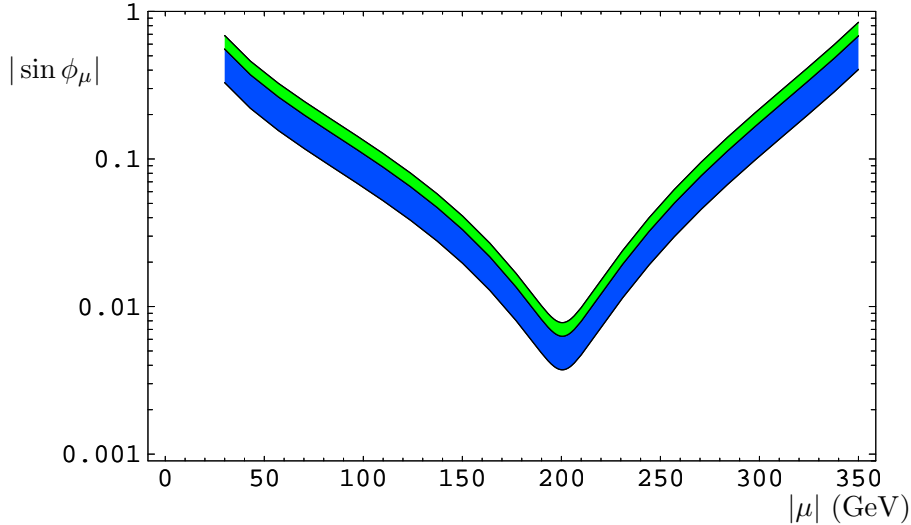,width=11cm}}}
\put(265,-9){{\small  
$\displaystyle |\mu| \ ({\rm GeV})$ }}
\put(-30,157){{\small 
$|\sin \phi_\mu| $ 
}}
\end{picture}
\caption{
Allowed band in the $ |\sin \phi_\mu| $--$|\mu|$ plane,
obtained by requiring successful electroweak baryogenesis.  All other 
MSSM parameters are given in the text.  The light-shaded (green) narrow band
corresponds to the experimental input from WMAP,  
while the two bands combined [dark (blue) + light (green)] correspond to input from 
Big Bang Nucleosynthesis. 
\label{fig:phimu}
}
\end{figure}

With the above choice of MSSM parameters, the stop-induced
contribution to $Y_B$ is suppressed ($F_2 \sim 10^{-3} F_1$),
since one is far off the squark resonance [$(M_{\tilde{t}_L} - M_{\tilde{t}_R}) \gg M_{\tilde{t}_R})$].
On the other hand, the Higgsino-induced contribution $F_1$ can account for
the observed $Y_B$ even without maximal values of $|\sin \phi_\mu |$.  
We highlight below the salient results of our study:
\begin{itemize}
\item The primary result of our analysis is that both the source 
$S_{\tilde{H}}^{\CPV}$ and the relaxation term $\Gamma_h$ display
the resonant behavior~\cite{Riotto:1998zb,Carena:1997gx} typical of
quantum transport for $|\mu| \sim M_2$. 
We illustrate this in Fig.~\ref{fig:sg}: the left panel shows the
behavior of the rescaled $CP$-violating higgsino source 
${\hat S}_{\tilde H} \equiv -S_{\tilde{H}}^{\CPV}/(v^2 \dot{\beta} \sin \phi_\mu)$
versus $|\mu|$, while the right panel displays the ratio $R_\Gamma$ of the
relaxation term $(\Gamma_h + \Gamma_{M}^{-})$ as calculated in this
work to the one used in previous studies, $(\Gamma_h + \Gamma_{M}^{-})_{H.N.}$~\cite{Huet:1995sh}.  To our
knowledge this is the first explicit calculation showing resonance
behavior for the relaxation term ${\bar\Gamma}\sim r_\Gamma (\Gamma_h + \Gamma_{M}^{-})$.

\item Since $F_1$ is proportional to
$S_{\tilde{H}}^{\CPV}/\sqrt{\Gamma_{h} + \Gamma_{M}^-}$,
the baryon asymmetry retains a resonant behavior, albeit with an
attenuation of the peak due to the enhanced relaxation term. This is
shown explicitly in Fig.~\ref{fig:F1}.  In the left panel we plot
$F_1/Y_{B}^{\rm WMAP}$, normalizing to the baryon asymmetry extracted
from CMB studies~\cite{wmap}: $Y_B^{\rm WMAP} = (9.2 \pm 1.1)\times
10^{-11}$ (the quoted error corresponds to $95 \%$ CL). 

\item For completeness we also display in Fig.~\ref{fig:F1} (right
panel) the behavior of the squark contribution $F_2/Y_{B}^{\rm WMAP}$
as a function of $M_{\tilde{t}_L}$, with $M_{\tilde{t}_R}= 100$ GeV.
Within the MSSM, precision electroweak data and the requirement that
$v(T_c)/T_c \gsim 1$ force the masses to be far away from the peak region.
However, in extensions of the MSSM where the phase transition is
strenghtened by additional scalar degrees of freedom this
contribution might be important (see, {\em e.g.}, Refs.~\cite{Dine:2003ax,Kang:2004pp}). 

\item For given values of the MSSM parameter space explored here, successful EWB carves out
a band in $|\sin \phi_\mu|$ centered at $|\sin \phi_\mu| = Y_{B}^{\rm
exp}/|F_1|$ (whose width depends on the uncertainty in $Y_B^{\rm
exp}$).  Due to the resonant behavior of $F_1$, the location of this
band is highly sensitive to the relative size of $M_2$ and $|\mu|$. As
illustration, in Fig.~\ref{fig:phimu} we plot the allowed band in the
$|\sin \phi_\mu|$--$|\mu|$ plane determined by the baryon asymmetry,
with all other MSSM parameters fixed as above.  The 
bands in the plot combined together correspond to the baryon density determined from Big
Bang Nucleosynthesis, $Y_B^{\rm BBN} = (7.3 \pm 2.5)\times 10^{-11}$
(the error corresponds to $95 \%$ CL~\cite{pdg04}).  Using WMAP
input leads to the narrow, lighter-shaded band in our plot located at the upper edge
of the BBN-induced band.
\end{itemize}

We conclude this section with a brief account of the connections
between the baryon asymmetry and EDM phenomenology. Since the Standard
Model predictions are in general highly suppressed and well below
present experimental sensitivity, limits on the electron, neutron, and
atomic EDMs can be used to constrain the phases of a given new physics
model.  Present limits of interest to us are:
\begin{equation*}
|d_e| < 1.9 \times 10^{-27} e \cdot \text{cm} 
\ (95 \%\text{ CL})~\text{\cite{Regan:2002ta}} \ , 
\qquad   \qquad 
|d_{Hg}| < 2.1 \times 10^{-28} e \cdot \text{cm}
\ (95 \%\text{ CL})~\text{\cite{Romalis:2000mg}}
\ .
\end{equation*}
Although a single EDM can be sufficiently small even for maximally
large $CP$-violating phases (due to cancellations), constraints from more than one
EDM can be very powerful.  In Ref.~\cite{Falk:1999tm}, for example, it was pointed
out how limits on electron and $^{199}$Hg EDMs single out a well
defined region in the $\phi_\mu$--$\phi_A$ plane, for given values of
gauginos, squark and slepton masses.  As shown above, for each point
in the MSSM paramter space, electroweak baryogenesis also selects a
band in the $\phi_\mu$--$\phi_A$ plane.  This implies in general
non-trivial constraints on the MSSM parameter space, as the EDM-allowed region
need not in general coincide with the one required by the baryon asymmetry.

\begin{figure}[!t]
\centering
\begin{picture}(300,250)  
\put(-10,100){\makebox(50,50){\epsfig{figure=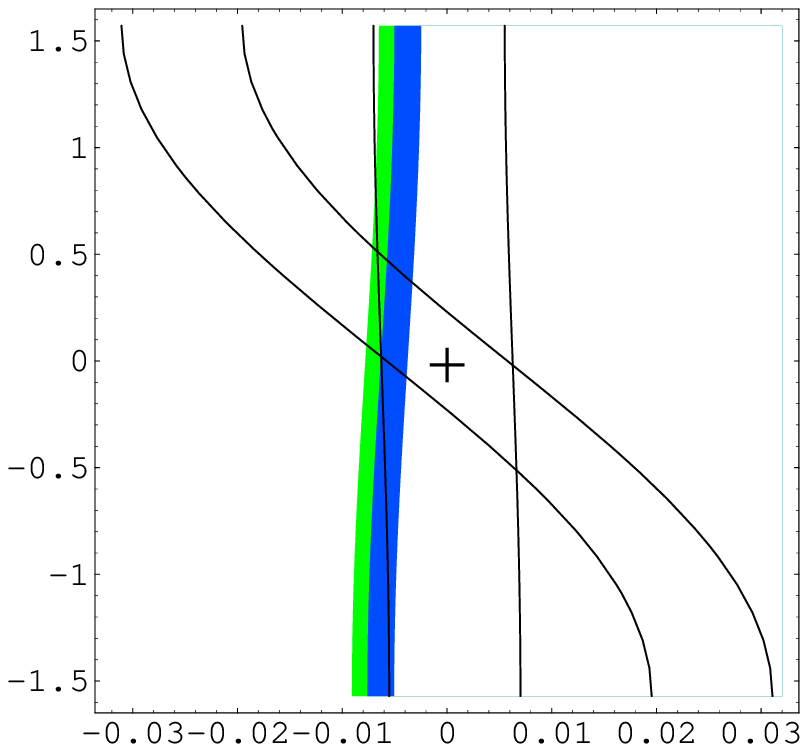,width=7.5cm}}}
\put(240,100){\makebox(50,50){\epsfig{figure=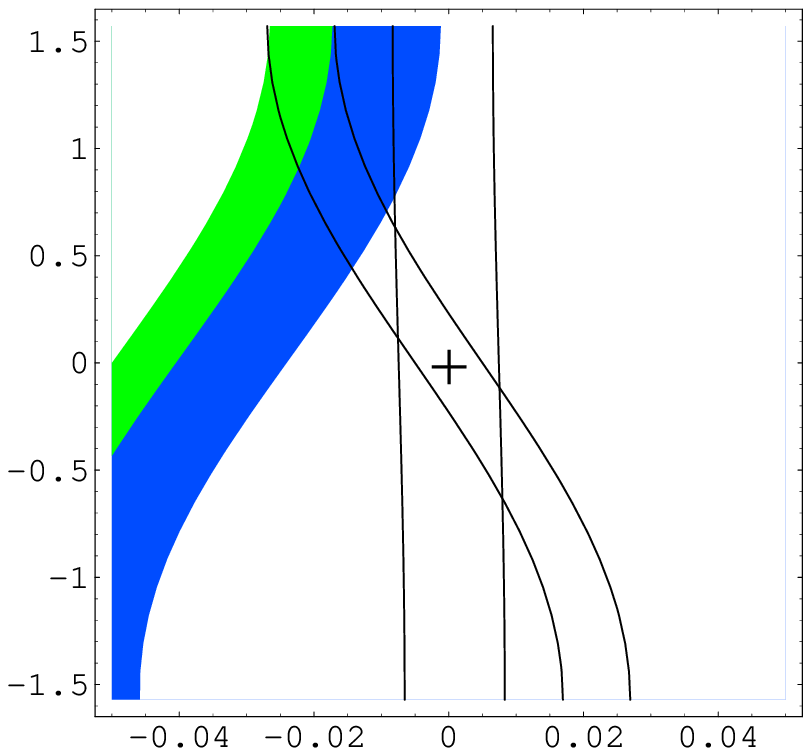,width=7.5cm}}}
\put(7,15){{\small $\phi_\mu$ (rad)}}
\put(260,15){{\small $\phi_\mu$ (rad)  }}
\put(-100,130){{\small $\phi_A$ }}
\put(-105,115){{\small (rad)}}
\put(150,130){{\small $\phi_A$ }}
\put(145,115){{\small (rad)}}
\put(250,70){{\small $d_e$ }}
\put(320,90){{\small $d_{Hg}$ }}
\put(205,75){{\scriptsize EWB}}
\put(-20,60){{\scriptsize EWB }}
\put(45,170){{\small $d_e$ }}
\put(80,100){{\small $d_{Hg}$ }}
\end{picture}
\caption{ Allowed bands in the $\phi_\mu$--$\phi_A$ plane implied by
consistency with the $95 \%$ C.L. limits on electron EDM ($|d_e| < 1.9
\times 10^{-27} e \cdot {\rm cm} $~\cite{Regan:2002ta}), 
mercury EDM ($|d_{Hg}| < 2.1 \times 10^{-28} e 
\cdot {\rm cm}$~\cite{Romalis:2000mg}), and baryogenesis.  
The shaded [dark (blue) and light (green) combined] EWB band corresponds to BBN
input~\cite{pdg04}, while the narrow light-shaded (green)
band on the left corresponds to WMAP input~\cite{wmap}. 
In the  left panel we use $|\mu| = M_2 = 200$ GeV (resonance peak), 
while in the right panel we use 
$M_2= 200$ GeV and $|\mu| = 250$ GeV (off resonance). In both cases 
the other supersymmetric masses are as specified in the text.  
\label{fig:bands}
}
\end{figure}

To illustrate this situation, we have evaluated the bands in $\phi_\mu-\phi_A$
allowed by present limits on electron EDM, mercury EDM, and EWB
for two representative points in the MSSM parameter space
(see Fig.~\ref{fig:bands}).  In our analysis we take the expressions
for the electron EDM and quark chromo-electric dipole moments from
Ref.~\cite{Ibrahim:1997gj}. In relating the $^{199}$Hg EDM to the
quark-level $CP$-violating couplings, we follow the treatment of
Ref.~\cite{Falk:1999tm}~\footnote{For a recent reanalysis of hadronic
EDMs in SUSY see Ref.~\cite{Hisano:2004tf}.}, where it was shown that
the dominant contribution arises from the chromo-electric dipole
moments of quarks ($\tilde{d}_q$) according to
\be
d_{Hg} = - \left(\tilde{d}_d - \tilde{d}_u - 0.012 \tilde{d}_s \right) 
\times 3.2 \cdot 10^{-2} e  \ . 
\ee
According to the same authors, the analysis of the neutron EDM involves additional effects, such as the $CP$-violating three-gluon operator ${\tilde G}GG$, that require a detailed analysis going beyond the scope of the present work. Although the experimental bounds on the neutron EDM---as well as the prospects for future improvements---are competitive with those for the electron and neutral atoms, we defer an analysis of its implications for EWB to a future study. 
We also neglect the renormalization group evolution of 
$\phi_\mu$ and $\phi_A$ from the weak scale to the atomic scale, having assumed a common, flavor-independent phase for the tri-scalar coupling at the former. 

The plots in Fig.~\ref{fig:bands} correspond to taking the first and
second generation sfermions, along with  the gluinos, to be  degenerate with masses equal to
750 GeV; the gaugino mass $M_1 = 100$ GeV; and the triscalar coupling
$A = 200$ GeV.  We consider then two cases for $M_2$ and $\mu$: the
left panel corresponds to the resonance peak $M_2=|\mu| = 200$ GeV,
while the right panel corresponds to off-resonance parameters $M_2=
200$ GeV, $|\mu| = 250$ GeV. For these choices of MSSM parameters, Eq.~(\ref{eq:pheno1}) predicts for $Y_B$:
\begin{align*}
M_2 = \abs{\mu} = 200\text{ GeV}:\qquad Y_B &= -1.3\times 10^{-8}\sin\phi_\mu + 1.7\times 10^{-11}\sin(\phi_A + \phi_\mu) \\
M_2 = 200\text{ GeV}, \abs{\mu} = 250\text{ GeV}:\, Y_B &= -2.0\times 10^{-9}\sin\phi_\mu + 4.6\times 10^{-11}\sin(\phi_A + \phi_\mu)
\end{align*}
These cases illustrate the main trend:
for $M_2 \sim |\mu|$ electroweak baryogenesis requires relatively
small phases, and is consistent with the constraints from EDMs.  As
one moves off resonance, then larger phases are needed to generate
the observed baryon asymmetry, and this requirement tends to conflict with the EDM constraints.  Indeed, within our simplified analysis, we find that
baryogenesis and EDM constraints become inconsistent for $|\mu| - M_2
\gsim 50$ GeV, when all other superpartners are kept around $750$ GeV.
Of course, increasing (decreasing) the sfermion masses relaxes
(tightens) the EDM limits on $CP$-violating phases and affects the above
conclusion. 

Ultimately, if supersymmetry is discovered at collider experiments,
spectroscopy will dictate the input for mass parameters. Then joint
constraints from low-energy EDM measurements and collider searches
could be used to tightly test the scenario of baryogenesis at the
electroweak scale.

\section{Conclusions}
\label{sec:summary}

It is instructive to consider the essential physics leading to the enhanced sources and relaxation terms discussed in this work. The propagation of quasiparticles in the plasma is modified by scattering from the spacetime  varying Higgs vevs that causes transitions to intermediate states involving other quasiparticle species. The system retains some memory of each scattering due to the presence of thermal widths, $\Gamma_i$, that reflect the degeneracy of states in the thermal bath. For $\Gamma_i=0$, the oscillating exponentials appearing in the Green's functions wash out any memory of the scattering. For $\Gamma_i\not\!= 0$, the Green's functions now contain decaying exponentials as well as oscillating terms, and the memory wash out is incomplete. The impact of quantum memory effects are, thus, characterized by the  ratio of time scales, $\tau_{\rm int}/\tau_p\sim \Gamma_i/\omega_i$, where $\tau_{\rm int}$ is the characteristic propagation time associated with a quasiparticle of frequency $\omega_i$ and $\tau_p\sim 1/\Gamma_i$, the plasma time, is time scale on which transitions between the quasiparticle and other, degenerate states may occur. To the extent that the quasiparticle thermal mass and/or three-momentum is large compared to $\Gamma_i$, this ratio $\tau_{\rm int}/\tau_p$ is ${\cal O}(\epsilon)$.

A special situation arises, however, when the spacetime variation of the Higgs vevs is gentle and the thermal mass of an intermediate state is close to that of the initial state. Under these conditions, the scattering event injects essentially zero four-momentum into the initial state $i$, leading to resonant production of the intermediate state $j$. The characteristic lifetime of the latter is no longer $\tau_{\rm int}\sim 1/\omega_i$, but rather the resonance time scale 
\be
\tau_{\rm res}\sim \frac{1}{\sqrt{\Delta\omega^2+\Gamma_{ij}^2}}\ \ \ ,
\ee
where $\Delta\omega=\omega_i-\omega_j$ and $\Gamma_{ij}=\Gamma_i+\Gamma_j$ [see, {\em e.g.}, Eqs. (\ref{appx:scalar5}-\ref{appx:gammascalarpm}) and (\ref{appx:gammaHiggsinopm}-\ref{appx:chargino3}) of Appendix \ref{appx:expand}]. In this case, the impact of quantum memory is characterized by the ratio $\tau_{\rm res}/\tau_p$. For $|\Delta\omega| \ll \Gamma_{ij}$, this ratio becomes of ${\cal O}(1)$, and the impact of quantum memory is resonantly enhanced\footnote{An examination of Eqs. (\ref{appx:scalar5}-\ref{appx:gammascalarpm}) and (\ref{appx:gammaHiggsinopm}-\ref{appx:chargino3}) of Appendix \ref{appx:expand}, {\em etc.} indicates the presence of an additional, dynamical enhancement factor $\sim\omega/\sqrt{\Delta\omega^2+\Gamma_{ij}^2}$ in the relevant integrals.}. On the other hand, for $|\Delta\omega|\gg\Gamma_{ij}$, the ratio is ${\cal O}(\epsilon)$ and one returns to the more generic conditions. 

In this study, we have shown how this effect can enhance both the particle number-changing relaxation terms as well as the $CP$-violating sources that enter the transport equations relevant to electroweak baryogenesis. Importantly, the effect of resonant relaxation tends to mitigate the impact of resonantly-enhanced sources, as both enhancements occur under the same conditions for the electroweak model parameters (in this case, those of the MSSM). We suspect that analogous resonant effects occur in other transport coefficients, such as the $\Gamma_Y$ Yukawa terms discussed above, but that the conditions on model parameters leading to enhancements---owing to simple kinematic considerations---will be different. It may be, for example, that the Yukawa interactions are no longer fast compared to the Higgs vev induced transitions when the latter are resonantly enhanced, and in this case, the solution to the differential equations will differ from the general structure obtained here and by other authors. This possibility is one that should be explored in future work.

Additional refinements of the present analysis are clearly in order, including some form of all-orders resummation of the Higgs vev insertions (possibly along the lines proposed in Refs. \cite{Carena:2000id,Prokopec}) 
and a treatment of the axial charge transport equations via Eq. (\ref{eq:fermion1b}). In principle, one would also like to study the density dependence of the thermal frequencies and widths, the impact of nonzero gaugino densities, variations in bubble wall geometry, and possibly higher-order effects in $\epsilon$, such as the departure of $\delta f$ of the thermal distribution functions from their equilibrium values. In short, it is apparent that EWB is not yet a solved problem, but rather one that calls for additional study.

Undertaking this effort will be important for electroweak phenomenology. As illustrated here as well as in other studies (\emph{e.g.}, \cite{Balazs:2004ae}), determining the viability of EWB within a given electroweak model involves a detailed interplay of collider phenomenology, precision electroweak data, EDM searches, and a careful treatment of the dynamics of the electroweak phase transition. In particular, in light of the open questions pertaining to the latter, it is too soon to draw definitive conclusions about the implications of the next generation of EDM searches for the baryon asymmetry. One hopes, however, that by the time these searches obtain their first results, the context for their theoretical interpretation will have been further clarified.

\acknowledgments The authors are grateful for the support from the
Institute for Nuclear Theory, where part of this work was completed.
This work was supported in part by U.S. Department of Energy contracts
DE-FG03-02ER41215 and DE-FG03-92ER40701, and by a National Science
Foundation Grant PHY00-71856. VC was supported by a Sherman Fairchild
Fellowship. CL was supported in part by a National Defense Science and
Engineering Graduate Fellowship.

\appendix

\section{Propagators at Finite Temperature and Density}
\label{appx:props}

In this section, we derive some useful properties of propagators at finite temperature and density, using derivations based on those for the case of finite temperature and zero density in Refs.~\cite{Weldon:1989ys,Weldon:1999th}.

\subsection{General Structure of Fermion Propagators}

We begin with the spectral function for fermions at temperature $T = 1/\beta$ in the presence of a chemical potential $\mu$:
\begin{equation}
\label{rhodef}
\rho_{\alpha\beta}(x) = \frac{1}{Z}\Tr\left[e^{-\beta(\op{H}-\mu\op{N})}\{\psi_\alpha(x),\bar\psi_\beta(0)\}\right],
\end{equation}
where $Z = \Tr[e^{-\beta(\op{H}-\mu\op{N})}]$. It is convenient to define the retarded and advanced propagators:
\begin{subequations}
\label{SRSAdef}
\begin{align}
S^R(x) &= \theta(x^0)\rho(x) \\
S^A(x) &= -\theta(x^0)\rho(x),
\end{align}
\end{subequations}
supressing spinor indices. The Fourier transforms of $S^{R,A}(x)$ and $\rho(x)$ are related by:
\begin{subequations}
\begin{align}
S^R(k^0,\vect{k}) &= i\int_{-\infty}^\infty\frac{d\omega}{2\pi}\frac{\rho(\omega,\vect{k})}{k^0 - \omega + i\epsilon} \\
S^A(k^0,\vect{k}) &= i\int_{-\infty}^\infty\frac{d\omega}{2\pi}\frac{\rho(\omega,\vect{k})}{k^0 - \omega - i\epsilon} \,.
\end{align}
\end{subequations}
It is possible to express the momentum-space spectral function in terms of a single product of $\psi_\alpha(x)$ and $\bar\psi_\beta(x)$ instead of the anticommutator in Eq.~(\ref{rhodef}), whose Fourier transform is:
\begin{equation}
\begin{split}
\rho_{\alpha\beta}(\omega,\vect{k}) &= \int d^4 x\,e^{i(\omega t - \vect{k}\cdot\vect{x})}\rho_{\alpha\beta}(t,\vect{x}) \\
&= \int d^4 x\,e^{i(\omega t - \vect{k}\cdot\vect{x})}\frac{1}{Z}\sum_n\bra{n}e^{-\beta(\op{H}-\mu\op{N})}\bigl[\psi_\alpha(x)\bar\psi_\beta(0) + \bar\psi_\beta(0)\psi_\alpha(x)\bigr]\ket{n}.
\end{split}
\end{equation}
Now insert a complete set of states between the fermion fields:
\begin{equation}
\label{complete}
\begin{split}
\rho(\omega,\vect{k}) = \int d^4 x\,e^{i(\omega t - \vect{k}\cdot\vect{x})}\frac{1}{Z}\sum_{n,j}\Bigl[&\bra{n}e^{-\beta(\op{H}-\mu\op{N})}\psi_\alpha(x)\ket{j}\bra{j}\bar\psi_\beta(0)\ket{n} \\
+ &\bra{n}e^{-\beta(\op{H}-\mu\op{N})}\bar\psi_\beta(0)\ket{j}\bra{j}\psi_\alpha(x)\ket{n}\Bigr].
\end{split}
\end{equation}
We can rewrite the second term by switching summation labels and translating $\psi_\alpha$ from $x$ to 0:
\begin{equation}
\begin{split}
\sum_{n,j}&\bra{n}e^{-\beta(\op{H}-\mu\op{N})}\bar\psi_\beta(0)\ket{j}\bra{j}\psi_\alpha(x)\ket{n} \\
&=\sum_{j,n}e^{i(E_n-E_j)t} e^{-i(\vect{k}_n-\vect{k}_j)\cdot\vect x}e^{-\beta E_j}e^{\beta\mu(N_n+1)}\bra{n}\psi_\alpha(0)\ket{j}\bra{j}\bar\psi_\beta(0)\ket{n},
\end{split}
\end{equation}
which after integrating in Eq.~(\ref{complete}), becomes
\begin{equation}
\frac{1}{Z}\sum_{j,n}(2\pi)^4\delta(\omega+E_n-E_j)\delta^3(\vect{k}+\vect{k}_n-\vect{k}_j)e^{-\beta(E_n+\omega)}e^{\beta\mu(N_n+1)}\bra{n}\psi_\alpha(0)\ket{j}\bra{j}\bar\psi_\beta(0)\ket{n},
\end{equation}
where we used the first delta function to replace $E_j$ with $E_n+\omega$ in the exponential $e^{-\beta E_j}$. This can now be written:
\begin{equation}
e^{-\beta(\omega-\mu)}\frac{1}{Z}\int d^4 x\,e^{i(\omega t - \vect{k}\cdot\vect{x})}\sum_{n,j}\bra{n}e^{-\beta(\op{H}-\mu\op{N})}\psi_{\alpha}(x)\ket{j}\bra{j}\bar\psi_\beta(0)\ket{n}
\end{equation}
which is $e^{-\beta(\omega-\mu)}$ times the first term of Eq.~(\ref{complete}), so we conclude:
\begin{equation}
\label{rhomom1}
\rho(\omega,\vect{k}) = \left[1 + e^{-\beta(\omega-\mu)}\right]\int d^4 x\,e^{i(\omega t-\vect{k}\cdot\vect{x})}\frac{1}{Z}\Tr\bigl[e^{-\beta(\op{H}-\mu\op{N})}\psi_\alpha(x)\bar\psi_\beta(0)\bigr].
\end{equation}
Similarly, we could have manipulated the first term of Eq.~(\ref{complete}) in the same way, and derived the companion relation:
\begin{equation}
\label{rhomom2}
\rho(\omega,\vect{k}) = \left[1 + e^{\beta(\omega-\mu)}\right]\int d^4 x\,
e^{i(\omega t-\vect{k}\cdot\vect{x})}\frac{1}{Z}
\Tr\bigl[e^{-\beta(\op{H}-\mu\op{N})}\bar\psi_\beta(0)\psi_\alpha(x)\bigr].
\end{equation}
The Green's functions $S^>(k^0,\vect{k})$ and $-S^<(k^0,\vect{k})$ appear on the right-hand sides of Eqs.~(\ref{rhomom1},\ref{rhomom2}), giving the relations:
\begin{subequations}
\label{S>S<fromrho}
\begin{align}
S^>(k^0,\vect{k}) &= [1 - n_F(k^0-\mu)]\rho(k^0,\vect{k}) \\
S^<(k^0,\vect{k}) &= -n_F(k^0-\mu)\rho(k^0,\vect{k}),
\end{align}
\end{subequations}
where $n_F(x) = 1/(1 + e^x)$.

The various Green's functions satisfy the identities:
\begin{subequations}
\begin{align}
S^t(x,y) &= S^R(x,y) + S^<(x,y) = S^A(x,y) + S^>(x,y) \\
S^{\bar t}(x,y) &= S^>(x,y) - S^R(x,y) = S^<(x,y) - S^A(x,y)\,,
\end{align}
\end{subequations}
which follow directly from the definitions in
Eqs.~(\ref{eq:Greens1},\ref{SRSAdef}). Thus, using
Eq.~(\ref{S>S<fromrho}), the time- and anti-time-ordered propagators
can be expressed in terms of the retarded and advanced propagators:
\begin{subequations}
\label{StfromSR}
\begin{align}
S^t(k^0,\vect{k}) &= [1 - n_F(k^0-\mu)]S^R(k^0,\vect{k}) + 
n_F(k^0-\mu)S^A(k^0,\vect{k}) \\
S^{\bar t}(k^0,\vect{k}) &= -n_F(k^0-\mu)S^R(k^0,\vect{k}) - 
[1-n_F(k^0-\mu)]S^A(k^0,\vect{k})\,.
\end{align}
\end{subequations}
Also note that $\rho = S^R-S^A = S^> - S^<$.

\subsection{Bosonic Propagators}

Similar results may be derived from scalar bosonic propagators, for which the analog to Eq.~(\ref{S>S<fromrho}) is:
\begin{subequations}
\begin{align}
G^>(k^0,\vect{k}) &= [1+n_B(k^0-\mu)]\rho(k^0,\vect{k}) \\
G^<(k^0,\vect{k}) &= n_B(k^0-\mu)\rho(k^0,\vect{k})\,,
\end{align}
\end{subequations}
where the momentum-space spectral function $\rho(k^0,\vect{k})$ for bosons is the Fourier transform of:
\begin{equation}
\rho(x) = \frac{1}{Z}\Tr\left\{e^{-\beta(\op{H}-\mu\op{N})}[\phi(x),\phi^*(0)]\right\}\,.
\end{equation}
The bosonic propagators also satisfy the identity $\rho = G^R - G^A = G^> - G^<$.

\subsection{Tree-Level Propagators}

At tree level, the propagators $S^{R,A}$ for fermions are given by:
\begin{equation}
S^{R,A}(k^0,\vect{k}) = \frac{i(\diracslash{k} + m)}{(k^0\pm i\epsilon)^2 - E_{\vect{k}}^2}\,,
\end{equation}
and $G^{R,A}$ for bosons are given by:
\begin{equation}
G^{R,A}(k^0,\vect{k}) = \frac{i}{(k^0\pm i\epsilon)^2 - E_{\vect{k}}^2}\,,
\end{equation}
where $E_{\vect{k}}^2 = \abs{\vect{k}}^2 + m^2$. Note that these propagators are independent of the temperature and chemical potential, which only enter in the thermal distribution functions appearing in the relations of the retarded and advanced propagators to the other Green's functions, for example, in Eq.~(\ref{StfromSR}).

\subsection{One-Loop Corrections to Massless Fermion Propagators}

Resumming the one-loop self-energy into the fermion propagator at finite temperature changes the pole structure of the propagator dramatically, introducing a new collective ``hole'' excitation of the plasma \cite{Klimov,Weldon:1989ys}. In fact, this structure can be shown to hold even beyond perturbation theory \cite{Weldon:1999th}. Extending the results of Ref.~\cite{Weldon:1999th} to include dependence on a chemical potential, the propagator takes the form given in Eqs.~(\ref{eq:slambdaint}--\ref{eq:rhominus}). Recall that in those equations $\mathcal{E}_{p,h} = \omega_{p,h} - i\Gamma_{p,h}$ are the complex poles of the spectral function, and $Z_{p,h}$ are the corresponding residues. At leading order in the ``hard thermal loop'' approximation (see Ref.~\cite{LeBellac}), calculating the poles only to order $\mathcal{E}\sim gT$, one finds $\Gamma = 0$, and $Z_{p,h}(k,\mu)$ and $\omega_{p,h}(k,\mu)$, where $k = \abs{\vect{k}}$, depend only quadratically on $\mu/T$, which we thus neglect in our analysis in the present work, where we keep only effects linear in $\mu/T$. In this limit, and including only a single gluon loop in the quark self-energy diagram, the poles of the spectral function are given by the solutions to the equation:
\begin{equation}
0 = k^0 - k - \frac{\alpha_s C_F\pi T^2}{4k}\left[\left(1-\frac{k^0}{k}\right)\log\abs{\frac{k^0+k}{k^0-k}} + 2\right]\,,
\end{equation}
where $C_F = 4/3$ is the Casimir of the fundamental representation of $SU(3)$. The solutions to this equation give the poles $k^0 = E_p(k), -E_h(k)$. The residues satisfy:
\begin{equation}
Z_{p,h}(k) = \frac{E_{p,h}^2 - k^2}{m_f^2}\,,
\end{equation}
where
\begin{equation}
m_f^2 = \frac{\alpha_s C_F\pi T^2}{2}\,.
\end{equation}
Calculation of the imaginary parts $\Gamma_{p,h}$ of the poles, since they begin at order $g^2 T$, requires a resummation of hard thermal loops in self-energy diagrams \cite{Braaten:1989mz,Braaten:1991gm,Braaten:1992gd}. We are also interested in their dependence on the chemical potential $\mu$. We leave the calculation of these effects to a future study.

\section{Expanded Source Terms for Quantum Transport}
\label{appx:expand}

\subsection{Bosons}

The $CP$-conserving source term for right-handed stops in Eq.~(\ref{eq:scalar5}) can be expanded by explicitly taking the imaginary part of the integrand:
\begin{align}
\label{appx:scalar5}
S^{CP}_{{\tilde t}_R}(x) = -\frac{1}{T}&\frac{N_C y_t^2}{2\pi^2}\abs{A_t v_u(x) - \mu^* v_d(x)}^2 
\int_0^\infty\frac{k^2 dk}{\omega_R\omega_L} \\
\nonumber
\times\biggl\{&\mu_R\left[{1\over\Delta}\left(\sin\phi\Imag h_R^++\cos\phi\
 {\rm Re}\ h_R^+\right)
-{1\over\delta}\left(\cos\theta\Real h_R^+-\sin\theta\Imag h_R^+\right)\right]\\
\nonumber
+&\mu_L\left[{1\over\Delta}\left(\sin\phi\Imag h_L^+-\cos\phi\Real h_L^+\right)
+{1\over\delta}\left(\cos\theta\Real h_L^+ - \sin\theta\Imag h_L^+\right)\right]\biggr\}\,,
\end{align}
where
\bea
\label{appx:scalardefs}
\omega_{L,R} & = & \sqrt{\abs{\vect{k}}^2+M_{{\tilde t}_{L,R}}^2}\\
\nonumber
\Delta & = & \sqrt{(\Gamma_L + \Gamma_R)^2+(\omega_L - \omega_R)^2}\\
\nonumber
\delta & = & \sqrt{(\Gamma_L + \Gamma_R)^2+(\omega_L + \omega_R)^2}\\
\nonumber
\tan\theta & = & \frac{\omega_L + \omega_R}{\Gamma_L + \Gamma_R} \\
\nonumber
\tan\phi & = & \frac{\omega_L - \omega_R}{\Gamma_L + \Gamma_R} \\
\nonumber
h_{L,R}^\pm & = & \frac{\exp[(\omega_{L,R}\pm i\Gamma_{L,R})/T]}
{\{\exp[(\omega_{L,R}\pm i\Gamma_{L,R})/T]-1\}^{2}}
\eea
and where $\Gamma_{L,R}$ are the thermal widths for the ${\tilde
t}_{L,R}$. The rates $\Gamma_{\tilde t}^\pm$ defined in Eq.~(\ref{eq:gammascalarpm}) can then be expressed:
\begin{align}
\label{appx:gammascalarpm}
\Gamma_{\tilde t}^{\pm} = -\frac{1}{T}\frac{y_t^2}{4\pi^2}\abs{A_t v_u(x) - \mu^* v_d(x)}^2 & \\
\nonumber
\times\int_0^\infty\frac{k^2 dk}{\omega_R\omega_L}
\biggl\{\frac{1}{\Delta}&\left[\sin\phi\Imag(h_L^+\pm h_R^+)-
\cos\phi\Real(h_L^+\mp h_R^+)\right]\\
\nonumber
+{1\over\delta}&\left[\cos\theta\Real(h_L^+\mp h_R^+) -
\sin\theta \Imag(h_L^+\mp h_R^+)\right]\biggr\}\,.
\end{align}
Meanwhile, the $CP$-violating source given in Eq.~(\ref{eq:scalarcp1}) can be expanded:
\begin{align}
\label{appx:scalarcp1}
S^{\CPV}_{{\tilde t}_R}(x) = N_C y_t^2&\Imag
(\mu A_t) v(x)^2{\dot\beta}(x) \int_0^\infty{k^2 dk\over 2\pi^2}{1
\over\omega_L\omega_R} \\
\nonumber
\times\biggl\{ &{1\over\delta^2}\left[\Real \left(1+n_R^++n_L^+\right)
\sin 2\theta+  
\Imag\left(n_R^++n_L^+\right)\cos 2\theta\right]\\
\nonumber
+&{1\over\Delta^2}\left[\Real\left(n_R^+-n_L^+\right)
\sin 2\phi - 
\Imag\left(n_R^++n_L^+\right)\cos 2\phi\right]\biggr\}\,,
\end{align}
where $n_{L,R}^\pm = n_B(\omega_{\tilde t_{L,R}} \pm \Gamma_{L,R})$.
Our result agrees with that of Ref.~\cite{Riotto:1998zb} except for a
different relative sign in front of the $\cos 2\phi$ term and the overall factor of $N_C$.

\subsection{Massive Fermions}

The $CP$-conserving rates for Higgsino-gaugino interactions given in Eq.~(\ref{eq:gammaHiggsinopm}) can be expanded:
\begin{align}
\label{appx:gammaHiggsinopm}
\Gamma_{\widetilde H^\pm}^{\pm} =  g_2^2 {v(x)^2}&\frac{1}{T} 
\int_0^\infty{k^2 dk\over 2\pi^2}\left({1\over\omh\omw}\right) \\
\nonumber
\times\Biggl(
&{1\over\Delta}\biggl\{
\left[\omh\omw+\gamh\gamw- k^2 + M_2\abs{\mu}\cos\theta_\mu\sin 2\beta(x)\right] \\
\nonumber
&\qquad\qquad\times\Bigl[\cos\phi\Real(\kwtilp \mp \khtilp) - \sin\phi\Imag(\kwtilp \pm \khtilp)\Bigr]\\
\nonumber
&\qquad+\left[\gamh\omw-\gamw\omh\right]\left[\sin\phi\Real(\kwtilp\mp\khtilp)
+\cos\phi\Imag(\kwtilp\pm\khtilp)\right]\biggr\}\\
\nonumber
+&
{1\over\delta}\biggl\{
\left[\omh\omw-\gamh\gamw+ k^2- M_2\abs{\mu}\cos\theta_\mu\sin 2\beta(x)\right] \\
\nonumber
&\qquad\qquad\times\Bigl[
\cos\theta\Real(\kwtilp\mp\khtilp) -\sin\theta\Imag(\kwtilp\mp\khtilp)\Bigr]\\
\nonumber
&\qquad-\left[\gamh\omw+\gamw\omh\right]\left[
\cos\theta\Imag(\kwtilp\mp\khtilp)+\sin\theta\Real(\kwtilp\mp\khtilp)\right]\biggr\}\Biggr)
\end{align}
where
\bea
\label{appx:fermiondefs}
\omega_{\widetilde H,\widetilde W} & = & \sqrt{\abs{\vect{k}}^2+M_{\widetilde H,\widetilde W}^2}\\
\nonumber
\Delta & = & \sqrt{(\Gamma_{\widetilde W} + \Gamma_{\widetilde H})^2+(\omega_{\widetilde W} - \omega_{\widetilde H})^2}\\
\nonumber
\delta & = & \sqrt{(\Gamma_{\widetilde W} + \Gamma_{\widetilde H})^2+(\omega_{\widetilde W} + \omega_{\widetilde H})^2}\\
\nonumber
\tan\theta & = & \frac{\omega_{\widetilde W} + \omega_{\widetilde H}}{\Gamma_{\widetilde W} + \Gamma_{\widetilde H}} \\
\nonumber
\tan\phi & = & \frac{\omega_{\widetilde W} - \omega_{\widetilde H}}{\Gamma_{\widetilde W} + \Gamma_{\widetilde H}} \\
\nonumber
h_{{\widetilde W},{\widetilde H}}^\pm &=& \frac{\exp[(\omega_{{\widetilde W},{\widetilde H}}\pm i\Gamma_{{\widetilde W},{\widetilde H}})/T]}{\{\exp[(\omega_{{\widetilde W},{\widetilde H}}\pm i\Gamma_{{\widetilde W},{\widetilde H}})/T]+1\}^{2}}
\end{eqnarray}
The $CP$-violating Higgsino source in Eq.~(\ref{eq:chargino3}) can be expressed:
\begin{align}
\label{appx:chargino3}
S^{\CPV}_{\widetilde H^\pm}(x)  =  
\nonumber
2 g_2^2 M_2 &\Imag(\mu) v(x)^2{\dot\beta}\ \int_0^\infty\ 
{k^2 dk\over 2\pi^2}\left({1\over \omh\omw}\right)
 \\
\times\biggl\{&{1\over\Delta^2}\left[\sin 2\phi\ {\rm Re}\left(\nwtilp-\nhtilp\right)+\cos 2\phi\ 
{\rm Im}\left(\nwtilp+\nhtilp\right)\right]\\
\nonumber
+&{1\over\delta^2}\left[\sin 2\theta\ {\rm Re}\left(1-\nwtilp-\nhtilp\right) 
-\cos 2\theta\ {\rm Im}\left(\nwtilp+\nhtilp\right)\right]\biggr\}\,,
\end{align}
where $N_{\widetilde H,\widetilde W}^\pm = n_B(\omega_{\widetilde H,\widetilde W}\pm i\Gamma_{\widetilde H,\widetilde W})$.
Our result agrees with that of Ref. \cite{Riotto:1998zb} except for
the sign of the $\cos 2\phi$ term.

\subsection{Chiral Fermions}
For chiral fermions, the $CP$-conserving chirality-changing rates in Eq.~(\ref{eq:gquarkplusminus}) can be expanded:
\begin{align}
\label{appx:gquarkplus}
\Gamma_{t_R}^{\pm} = \frac{1}{T}\frac{N_C y_t v_u^2}{\pi^2}&\int_0^\infty k^2 dk \\
\times\biggl\{\frac{Z_p^R Z_p^L}{\delta_p}&\Bigr[\sin\theta_p\bigl\{\Real(\lambda_p^L \kplp \mp \lambda_p^R \kprp) - \Imag(\kplp\mp \kprp)\bigr\} + \cos\theta_p\Real(\kplp \mp \kprp) \nonumber \\
 &+ \frac{T}{\delta_p}\cos 2\theta_p(\lambda_p^L \mp \lambda_p^R)\Real(1-N_{pL}^+ - N_{pR}^+)\Bigr] \nonumber \\
-\frac{Z_p^L Z_h^R}{\Delta_{hp}}&\Bigl[\sin\phi_{hp}\bigl\{\Real(\lambda_p^L\kplp \pm \lambda_h^R\khrp) - \Imag(\kplp \pm \khrp)\bigr\} - \cos\phi_{hp}\Real(\kplp \mp \khrp) \nonumber \\
& + \frac{T}{\Delta_{hp}}\cos 2\phi_{hp}(\lambda_p^L\pm\lambda_h^R)\Real(N_{pL}^+ - N_{hR}^+)\Bigr] \nonumber \\
+ (p\leftrightarrow h) &\biggr\} \nonumber
\end{align}
where
\begin{equation}
\begin{split}
\delta_p &= \sqrt{(\omega_p^R + \omega_p^L)^2 + (\Gamma_p^R + \Gamma_p^L)^2} \\
\Delta_{hp} &= \sqrt{(\omega_p^L - \omega_h^R)^2 + (\Gamma_h^R + \Gamma_p^L)^2} \\
h_{pL}^\pm &= h_F(\omega_p^L \pm i\Gamma_p^L)\text{, etc.} \\
N_{pL}^\pm &= n_F(\omega_p^L \pm i\Gamma_p^L)\text{, etc.} \\
\tan\theta_p &= \frac{\omega_p^L + \omega_p^R}{\Gamma_p^L + \Gamma_p^R} \\
\tan\phi_{hp} &= \frac{\omega_h^R - \omega_p^L}{\Gamma_p^L + \Gamma_h^R},
\end{split}
\end{equation}
and where the
\begin{equation}
\lambda_{p,h}^{L,R} = \pd{\Gamma_{p,h}^{L,R}}{\mu_{t_{L,R}}}\,,
\end{equation}
parameterize the linear shifts in the thermal widths due to non-vanishing chemical potential. As noted at the end of Appendix~\ref{appx:props}, in a fully resummed calculation of the fermion self-energy, such shifts which are linear in $\mu_i/T$ may arise, and thus have to be included in our calculations, which we defer to future work. Also note that we have approximated the residues $Z_{p,h}^{L,R}$ to be purely real, which is true at the order we are working.

\end{document}